\documentclass[aps,prd,preprintnumbers,nofootinbib,showpacs]{revtex4}
\usepackage{graphicx}
\usepackage{amsfonts}
\usepackage{amssymb,amsmath}
\usepackage{color}
\usepackage[colorlinks=true,linkcolor=blue,citecolor=blue]{hyperref}

\newcommand{\simgt}{\lower.5ex\hbox{$\; \buildrel > \over \sim \;$}}
\newcommand{\simlt}{\lower.5ex\hbox{$\; \buildrel < \over \sim \;$}}

\begin{document}

\begin{center}

{\bf  
Third order solutions of the cosmological density perturbations
in the Horndeski's most general scalar-tensor theory with the 
Vainshtein mechanism}

\vskip .45in

{
Yuichiro Takushima, Ayumu Terukina, Kazuhiro Yamamoto
}

\vskip .45in

{\em
Department of Physical Science, Graduate School of Science, Hiroshima University, \\
Higashi-hiroshima, 739-8526, Japan
}

\begin{abstract}
We study the third order solutions of the cosmological density perturbations
in the Horndeski's most general scalar-tensor theory under the condition
that the Vainshtein mechanism is at work.  
In this work, we thoroughly investigate the independence property of 
the functions describing the nonlinear mode-couplings, which is also useful 
for models within the general relativity. 
Then, we find that the solutions of the density contrast and
the velocity divergence up to the third order ones are characterized by 
6 parameters. Furthermore, the 1-loop order power spectra obtained with 
the third order solutions are described by 4 parameters. 
We exemplify the behavior of
the 1-loop order power spectra assuming the kinetic gravity 
braiding model, which demonstrates that the effect of the modified 
gravity appears more significantly in the 
power spectrum of the velocity divergence than the density contrast. 
\end{abstract}

\end{center}
\vskip .4in


\maketitle

\setcounter{page}{1}
\section{Introduction} 
\label{sec:intro}

The accelerated expansion of the Universe is one of the most 
fundamental problems in modern cosmology. The standard cosmological 
model introducing the cosmological constant is consistent with various 
observations \cite{Peebles,PlanckI}. However, the small value of the cosmological 
constant raises the problem of fine tuning  \cite{Weinberg,Weinberg2,Martin}. 
As an alternative to the cosmological constant, the cosmic accelerated 
expansion might be explained by modifying gravity theory, e.g., 
\cite{HuSawicki,Starobinsky,Tsujikawafr,Nojirib,DGP2,Song,Maartens,KM,MassiveG,MassiveG2,GCMG,MassiveG3,HA}. 
In the present paper, we focus on the most general scalar-tensor theory
with the second order differential field equations \cite{DGSZ,Koba}, 
which was first discovered by Horndeski \cite{Horndeski}.
The Horndeski's most general scalar-tensor theory, including 4 
arbitrary functions of the scalar field and kinetic term, 
reduces to various modified gravity models by choosing the specific 
4 functions. Because the Horndeski's theory includes a wide class of modified 
gravity models, we adopt it as an effective theory of the 
generalized theories of gravity. 

In the present paper, we investigate the aspects of the quasi-nonlinear
evolution of the cosmological density perturbations in the 
Horndeski's most general scalar-tensor theory assuming that 
the Vainshtein mechanism is at work \cite{Vainshtein,KKY,Kase,Narikawa}. 
The  Vainshtein mechanism is the screening mechanisms, which is 
useful to evade the constraints from the gravity tests in the solar system. 
We investigate the effects of the nonlinear terms in the matter's fluid equations
as well as the nonlinear derivative interaction terms in the scalar field equation. 
In a previous work \cite{TTY}, the second order solution of the cosmological 
density perturbations is obtained. In the present paper, we extend the 
analysis to the third order solution, which enables us to compute the 1-loop 
order matter power spectrum. 

There are many works on the higher order cosmological density 
perturbations and the quasi-nonlinear matter power spectrum, 
which have been developed from the standard perturbative approach
(see e.g., \cite{Juszkiewicz,Vishniac,Goroff,SutoSasaki,Makino,Bertshinger,
Scoccimarro,Bernardeau,Takahashi,Komatsu,Koyama}). 
Improvements to include the non-perturbative effects have been investigated, e.g., 
\cite{ScoccimarroM1,ScoccimarroM2,ScoccimarroM3,MatsubaraM,TaruyaM,taruya}, 
but we here adopt the standard perturbative approach of the cosmological density 
perturbations as a starting place for the analysis of the Horndeski's most general 
scalar-tensor theory. 
Related to the work of the present paper, we refer the recent work by Lee, 
Park and Biern \cite{fulltime}, in which a similar solution is obtained 
for the dark energy model within the general relativity. 

This paper is organized as follows. In section 2, we review the basic equations
and the second order solution \cite{TTY}. 
In section 3, we construct the third order solutions
of the cosmological density perturbations. Here, we carefully investigate independent
functions of mode-couplings describing nonlinear interactions.
In section 4, we derive the expression
of the 1-loop order power spectra of the matter density contrast and the velocity divergence. 
In section 5, an expression for the trispectrum for the density contrast is 
presented.
In section 6, 
we demonstrate the behavior of the 1-loop order power spectra in the kinetic 
gravity braiding model. Section 7 is devoted to summary and conclusions.
In appendix A, definitions of the coefficients to characterize the Horndeski's theory
are summarized. In appendix B, definitions of the functions to describe 
the nonlinear mode-coupling for the third order solutions are summarized. 
In appendix C, a derivation of the 1-loop power spectra is summarized. 
Expressions in appendix D are useful for the deviation of the 1-loop power 
spectra. Appendix E lists the coefficients to characterize the kinetic 
gravity braiding model.

\section{Review of the second order solution}
Let us start with reviewing the basic formulas \cite{KKY,TTY}.
We consider the Horndeski's most general scalar-tensor theory, 
whose action is given by
\begin{eqnarray}
S=\int d^4x\sqrt{-g}\left({\cal L}_{\rm GG}+{\cal L}_{\rm m}\right),\label{action}
\label{action}
\end{eqnarray}
where we define 
\begin{eqnarray}
{\cal L}_{\rm GG} &=& K(\phi, X)-G_3(\phi, X)\Box\phi
+G_4(\phi, X)R+G_{4X}\left[(\Box\phi)^2-(\nabla_\mu\nabla_\nu\phi)^2\right]
\nonumber\\
&&+G_5(\phi, X)G_{\mu\nu}\nabla^\mu\nabla^\nu\phi
-\frac{1}{6}G_{5X}\bigl[(\Box\phi)^3
-3\Box\phi(\nabla_\mu\nabla_\nu\phi)^2+
2(\nabla_\mu\nabla_\nu\phi)^3\bigr],\label{GG}
\end{eqnarray}
where $K, G_3, G_4,$ and $G_5$ are arbitrary function of the scalar field $\phi$ 
and the kinetic term $X:=-(\partial\phi)^2/2$, $G_{iX}$ denotes $\partial G_i/\partial X$, 
$R$ is the Ricci scalar, $G_{\mu\nu}$ is the Einstein tensor, and 
${\cal L}_{\rm m}$ is the Lagrangian of the matter field, 
which is minimally coupled to the gravity.

The basic equations for the cosmological density perturbations are 
derived in Ref.~\cite{KKY}. Here, we briefly review the method and 
the results (see \cite{KKY} for details). 
This theory is discovered in \cite{DGSZ} as a generalization of the galileon theory 
(\cite{Nicolis}, see also \cite{Koba,GC,SAUGC,ELCP,CEGH,GGC,GBDT,DPGMGT,CG,GGIR,MGALG,CCGF,OCG,CSTOG,FeliceTsujikawa,Deffayet,DDEF,RFFGM,SSSDBI,DBIGR}), 
but the equivalence with the Horndeski's theory~\cite{Horndeski} is shown in \cite{Koba}.
We consider a spatially flat expanding universe and the metric 
perturbations in the Newtonian gauge, whose line element is written as
\begin{eqnarray}
ds^2=-(1+2\Phi(t, \mathbf{x}))dt^2+a^2(t)(1-2\Psi(t, \mathbf{x}))d\mathbf{x}^2.
\end{eqnarray}
We define the scalar field with perturbations by 
\begin{eqnarray}
\phi&\to&\phi(t)+\delta\phi(t, \mathbf{x}),
\end{eqnarray}
and we introduce $Q=H{\delta\phi}/{\dot\phi}$.

The basic equations of the gravitational and scalar fields are derived 
on the basis of the quasi-static approximation
of the subhorizon scales \cite{KKY}. 
In the models that the Vainshtein mechanism may work, the basic equations 
can be found by keeping the leading terms schematically written as $(\partial\partial Y)^n$, 
with $n\geq1$, where $\partial$ denotes a spatial derivative and $Y$
does any of $\Phi$, $\Psi$ or $Q$.
Such terms make a leading contribution of the order $(L_{\rm H}^2\partial\partial Y)^n$, 
where $L_{\rm H}$ is a typical horizon length scale, and we have
\begin{eqnarray}
&&\nabla^2\left({\cal F}_T\Psi-{\cal G}_T\Phi-A_1 Q\right)
=\frac{B_1}{2a^2H^2}{\cal Q}^{(2)}
+\frac{B_3}{a^2H^2}\left(
\nabla^2\Phi\nabla^2Q-\partial_i\partial_j\Phi\partial^i\partial^j Q
\right),
\label{trlseq}
\end{eqnarray}
and
\begin{eqnarray}
{\cal G}_T\nabla^2\Psi
&=&\frac{a^2}{2}\rho_{\rm m}\delta
-A_2 \nabla^2 Q
-\frac{B_2}{2a^2H^2} {\cal Q}^{(2)}
\nonumber\\&&
-\frac{B_3}{a^2H^2}\left(\nabla^2\Psi\nabla^2Q
-\partial_i\partial_j\Psi\partial^i\partial^jQ\right)
-\frac{C_1}{3a^4H^4}{\cal Q}^{(3)},
\label{00eq}
\end{eqnarray}
where $\rho_{\rm m}$ is the background 
matter density and $\delta$ is the matter density contrast, and we define
\begin{eqnarray}
&&{\cal Q}^{(2)}:=\left(\nabla^2Q\right)^2-\left(\partial_i\partial_j Q\right)^2,
\\
&&{\cal Q}^{(3)}:=\left(\nabla^2 Q\right)^3
-3\nabla^2Q\left(\partial_i\partial_jQ\right)^2
+2\left(\partial_i\partial_jQ\right)^3.
\end{eqnarray}
The equation of the scalar field perturbation is 
\begin{eqnarray}
&&A_0\nabla^2Q
-A_1\nabla^2\Psi
-A_2\nabla^2\Phi+\frac{B_0}{a^2H^2}{\cal Q}^{(2)}
-\frac{B_1}{a^2H^2}
\left(\nabla^2\Psi\nabla^2Q-\partial_i\partial_j\Psi\partial^i\partial^jQ\right)
\nonumber\\&&
-\frac{B_2}{a^2H^2}
\left(\nabla^2\Phi\nabla^2Q-\partial_i\partial_j\Phi\partial^i\partial^jQ\right)
-\frac{B_3}{a^2H^2}
\left(\nabla^2\Phi\nabla^2\Psi -
\partial_i\partial_j\Phi\partial^i\partial^j\Psi \right)
\nonumber\\&&
-\frac{C_0}{a^4H^4}{\cal Q}^{(3)}
-\frac{C_1}{a^4H^4}{\cal U}^{(3)}= 0,
\label{seom}
\end{eqnarray}
where we define
\begin{eqnarray}
{\cal U}^{(3)}&:=&
{\cal Q}^{(2)}\nabla^2\Phi
-2\nabla^2Q\partial_i\partial_jQ\partial^i\partial^j\Phi
+2\partial_i\partial_jQ\partial^j\partial^kQ\partial_k\partial^i\Phi.
\end{eqnarray}
Here the coefficients ${\cal F}_T$, $A_1$, $B_1$, $C_1$, etc., are defined in Appendix A. 
$A_i$, $B_i$, and $C_i$ are the coefficients of the linear, quadratic and cubic terms of
$\Psi$, $\Phi$, and $Q$, respectively. 

From the continuity equation and the Euler equation for the matter fluid,
we have the following equations for the density contrast $\delta$ and the 
velocity field $u^i$,
\begin{eqnarray}
&&{\partial \delta(t,{\bf x})\over \partial t}
+{1\over a}\partial_i[(1+\delta(t,{\bf x})) u^i(t,{\bf x})]=0,
\label{continue}
\\
&&{\partial u^i(t,{\bf x})
\over \partial t}+{\dot a\over a}u^i(t,{\bf x})
+{1\over a}u^j(t,{\bf x})\partial_ju^i(t,{\bf x})
=-{1\over a}\partial_i\Phi(t,{\bf x}),
\label{Euler}
\end{eqnarray}
respectively. The properties of the gravity sector is influenced through 
$\Phi$ in (\ref{Euler}), where $\Phi$ is determined by Eqs.~(\ref{trlseq}), 
(\ref{00eq}) and (\ref{seom}).

Now introducing the scalar function $\theta\equiv \nabla {\bf u}/(aH)$, which we call 
velocity divergence, and we perform the Fourier expansions for $\delta$ and $\theta$,
\begin{eqnarray}
&&\delta(t,{\bf x})={1\over (2\pi)^3}\int d^3p\delta(t,{\bf p})e^{i{\bf p}\cdot{\bf x}},
\label{deftildedelta}
\\
&&u^j(t,{\bf x})={1\over (2\pi)^3}\int d^3p {-ip^j \over p^2}aH\theta(t,{\bf p})e^{i{\bf p}\cdot{\bf x}}.
\end{eqnarray}
In the similar way, we perform the Fourier expansions for 
$\Phi$, $\Psi$, and $Q$. Then, the equations for the gravity
(\ref{trlseq}) and (\ref{00eq}) lead to
\begin{eqnarray}
&&-p^2\left({\cal F}_T\Psi(t,{\bf p})-{\cal G}_T\Phi(t,{\bf p})-A_1 Q(t,{\bf p})\right)
=\frac{B_1}{2a^2H^2}\Gamma[t,{\bf p};Q,Q]
+\frac{B_3}{a^2H^2}\Gamma[t,{\bf p};Q,\Phi],
\label{se14}\\
&&-p^2({\cal G}_T\Psi(t,{\bf p})+A_2 Q(t,{\bf p}))
-\frac{a^2}{2}\rho_{\rm m}\delta(t,{\bf p})
\nonumber\\
&&~~~~~~~~=-\frac{B_2}{2a^2H^2} \Gamma[t,{\bf p};Q,Q]
-\frac{B_3}{a^2H^2}\Gamma[t,{\bf p};Q,\Psi]
-\frac{C_1}{3a^4H^4}
\Xi_1[t,{\bf p};Q,Q,Q],
\label{se15}
\end{eqnarray}
respectively, where we define
\begin{eqnarray}
&&\Gamma[t,{\bf p};Z_1,Z_2]={1\over (2\pi)^3}\int d{\bf k}_1 d{\bf k}_2\delta^{(3)}({\bf k}_1+{\bf k}_2-{\bf p})
\left(k_1^2k_2^2-({\bf k}_1\cdot{\bf k}_2)^2\right)Z_1(t,{\bf k}_1)Z_2(t,{\bf k}_2),\\
&&\Xi_1[t,{\bf p}; Z_1,Z_2,Z_3]={1\over (2\pi)^6}\int d{\bf k}_1 d{\bf k}_2d{\bf k}_3\delta^{(3)}({\bf k}_1+{\bf k}_2+{\bf k}_3-{\bf p})
\nonumber\\&&
\hspace{0cm}
~~~~~~~~~~~\times \biggl[
-k_1^2k_2^2k_3^2+3k_1^2({\bf k}_2\cdot{\bf k}_3)^2-2({\bf k}_1\cdot{\bf k}_2)({\bf k}_2\cdot{\bf k}_3)({\bf k}_3\cdot{\bf k}_1)\biggr]Z_1(t,{\bf k}_1)Z_2(t,{\bf k}_2)Z_3(t,{\bf k}_3),
\end{eqnarray}
where $Z_1$, $Z_2$ and $Z_3$ denote any of $Q$, $\Phi$ or $\Psi$. 
The equation for scalar field perturbation (\ref{seom}) leads to
\begin{eqnarray}
&&-p^2(A_0Q(t,{\bf p})
-A_1\Psi(t,{\bf p})
-A_2\Phi(t,{\bf p}))
\nonumber\\
&&~~~~=-\frac{B_0}{a^2H^2}\Gamma[t,{\bf p};Q,Q]
+\frac{B_1}{a^2H^2}\Gamma[t,{\bf p};Q,\Psi]
+\frac{B_2}{a^2H^2}\Gamma[t,{\bf p};Q,\Phi]
+\frac{B_3}{a^2H^2}\Gamma[t,{\bf p};\Psi,\Phi]
\nonumber\\
&&~~~~~~~~
+\frac{C_0}{a^4H^4}
\Xi_1[t,{\bf p};Q,Q,Q]
+\frac{C_1}{a^4H^4}
\Xi_2[t,{\bf p};Q,Q,\Phi],
\label{seomf}
\end{eqnarray}
where we define
\begin{eqnarray}
&&\Xi_2[t,{\bf p};Z_1,Z_2,Z_3]={1\over (2\pi)^6}\int d{\bf k}_1 d{\bf k}_2d{\bf k}_3\delta^{(3)}({\bf k}_1+{\bf k}_2+{\bf k}_3-{\bf p})
\biggl[
-k_1^2k_2^2k_3^2+({\bf k}_1\cdot{\bf k}_2)^2k_3^2
\nonumber\\&&
~~~~~~~~\hspace{0.1cm}
+2k_1^2({\bf k}_2\cdot{\bf k}_3)^2
-2({\bf k}_1\cdot{\bf k}_2)({\bf k}_2\cdot{\bf k}_3)({\bf k}_3\cdot{\bf k}_1)\biggr]
Z_1(t,{\bf k}_1)Z_2(t,{\bf k}_2)Z_3(t,{\bf k}_3).
\end{eqnarray}
The fluid equations (\ref{continue}) and (\ref{Euler}) lead to
\begin{eqnarray}
&&{1\over H}{\partial \delta(t,{\bf p})\over \partial t}+\theta(t,{\bf p})
=-{1\over (2\pi)^3}\int d{\bf k}_1 d{\bf k}_2\delta^{(3)}({\bf k}_1+{\bf k}_2-{\bf p})
\alpha({\bf k}_1,{\bf k}_2)\theta(t,{\bf k}_1)\delta(t,{\bf k}_2),
\nonumber\\
\label{se11}
\\
&&{1\over H}{\partial \theta(t,{\bf p})\over \partial t}+
\left(2+{\dot H\over H^2}\right)\theta(t,{\bf p})-{p^2\over a^2H^2}\Phi(t,{\bf p})
\nonumber\\
&&\hspace{3.0cm}
=-{1\over (2\pi)^3}\int d{\bf k}_1 d{\bf k}_2\delta^{(3)}({\bf k}_1+{\bf k}_2-{\bf p})
\beta({\bf k}_1,{\bf k}_2)
\theta(t,{\bf k}_1)\theta(t,{\bf k}_2),
\label{se12}
\end{eqnarray}
where we define 
\begin{eqnarray}
\alpha({\bf k}_1,{\bf k}_2) &=& 1+{{\bf k}_1\cdot{\bf k}_2\over k_1^2}\\
\beta({\bf k}_1,{\bf k}_2) &=& {({\bf k}_1\cdot {\bf k}_2)|{\bf k}_1+{\bf k}_2|^2\over 2 k_1^2k_2^2}.
\end{eqnarray}

Note that $\alpha({\bf k}_1, {\bf k}_2)$ doesn't have the symmetry with respect to exchange between ${\bf k}_1$ and ${\bf k}_2$.
We find the solution in terms of a perturbative expansion, which can 
be written in the form
\begin{eqnarray}
Y(t,{\bf p}) &=& \sum_{n=1}Y_n(t,{\bf p}),
\end{eqnarray}
where $Y$ denotes $\delta,~\theta,~\Psi,~\Phi$ or $Q$, and 
$Y_n$ denotes the $n$th order solution of the perturbative expansion. 
Neglecting the decaying mode solution, the linear order solution is written as 
\cite{DKT,KKY} 
\begin{eqnarray}
\delta_1(t,{\bf p})&=&D_+(t)\delta_{\rm L}({\bf p}),\\
\theta_1(t,{\bf p})&=& - D_+(t) f(t) \delta_{\rm L}({\bf p}),\\
\Phi_1(t,{\bf p}) &=& - {a^2 H^2 \over p^2} D_+(t)\kappa_\Phi(t)\delta_{\rm L}({\bf p}),\\
\Psi_1(t,{\bf p}) &=& -{a^2 H^2 \over p^2} D_+(t)\kappa_\Psi(t)\delta_{\rm L}({\bf p}),\\
Q_1(t,{\bf p}) &=& - {a^2 H^2 \over p^2} D_+(t)\kappa_Q(t)\delta_{\rm L}({\bf p}),
\end{eqnarray}
where $D_+(t)$ is growth factor obeying
\begin{eqnarray}
{d^2 D_+(t)\over dt^2} + 2 H {d D_+(t)\over dt} + L(t) D_+(t) = 0 \label{DEofd},
\label{linearDP}
\end{eqnarray}
with
\begin{eqnarray}
L(t) &=&  - {(A_0 {\cal F}_T- A_1^2) \rho_{\rm m}\over 2 (A_0 {\cal G}_T^2 + 2A_1A_2{\cal G}_T + A_2 {\cal F}_T)}
,\label{elut}
\end{eqnarray}
and $\delta_{\rm L}({\bf p})$ describes the linear density perturbations, which
are assumed to obey the Gaussian random distribution. 
Here we adopt the normalization for the growth factor 
$D_+(a) = a$ at $a \ll 1$, and introduced the 
linear growth rate defined by $f(t)=d\ln D_+(t)/\ln a$. 

The second order solution is written as (see \cite{TTY} for details),
\begin{eqnarray}
\delta_2(t,{\bf p})&=&D_+^2(t)\left({\cal W}_\alpha({\bf p}) - {2\over7} \lambda(t){\cal W}_\gamma({\bf p})\right),\\
\theta_2(t,{\bf p})&=& - D_+^2(t) f \left({\cal W}_\alpha({\bf p}) - {4\over7} \lambda_{\theta}(t){\cal W}_\gamma({\bf p})\right),\\
\Phi_2(t,{\bf p}) &=& - {a^2 H^2 \over p^2} 
D_+^2(t)(\kappa_\Phi(t){\cal W}_\alpha({\bf p}) + \lambda_\Phi(t){\cal W}_\gamma({\bf p})),\\
\Psi_2(t,{\bf p}) &=& - {a^2 H^2 \over p^2}
D_+^2(t)(\kappa_\Psi(t){\cal W}_\alpha({\bf p}) + \lambda_\Psi(t){\cal W}_\gamma({\bf p})),\\
Q_2(t,{\bf p}) &=& - {a^2 H^2 \over p^2}
D_+^2(t)(\kappa_Q(t){\cal W}_\alpha({\bf p}) + \lambda_Q(t){\cal W}_\gamma({\bf p})),
\end{eqnarray}
where the coefficients $\kappa_{\Phi}$, $\kappa_{\Psi}$, $\kappa_Q$, $\lambda$, 
$\lambda_\theta$, $\lambda_\Phi$, $\lambda_\Psi$, and $\lambda_Q$, 
are determined by the functions in the Lagrangian and the 
Hubble parameter, whose definitions are summarized in appendix A. 
Here ${\cal W}_\alpha({\bf p})$ and ${\cal W}_\gamma({\bf p})$ are defined as
\begin{eqnarray}
{\cal W}_\alpha({\bf p}) &=& {1\over (2 \pi)^3}\int d{\bf k}_1 d{\bf k}_2 \delta^{(3)}({\bf k}_1 +{\bf k}_2 - {\bf p})\alpha^{(s)}({\bf k}_1,{\bf k}_2)\delta_{\rm L}({\bf k}_1)\delta_{\rm L}({\bf k}_2),\\
{\cal W}_\gamma({\bf p}) &=& {1\over (2 \pi)^3}\int d{\bf k}_1 d{\bf k}_2 \delta^{(3)}({\bf k}_1 +{\bf k}_2 - {\bf p})\gamma({\bf k}_1,{\bf k}_2)\delta_{\rm L}({\bf k}_1)\delta_{\rm L}({\bf k}_2)
\end{eqnarray}
with 
\begin{eqnarray}
\alpha^{(s)}({\bf k}_1,{\bf k}_2) &=& 1 + {{\bf k}_1\cdot{\bf k}_2 (k_1^2 + k_2^2)\over2 k_1^2 k_2^2},\\
\gamma({\bf k}_1,{\bf k}_2) &=& 1 - {({\bf k}_1\cdot{\bf k}_2)^2\over k_1^2 k_2^2},
\end{eqnarray}
where $\alpha^{(s)}({\bf k}_1, {\bf k}_2)$ is obtained by symmetrizing 
$\alpha({\bf k}_1, {\bf k}_2)$ with respect to ${\bf k}_1$ and ${\bf k}_2$, 
and $\gamma({\bf k}_1, {\bf k}_2)$ is the function to describe the mode-couplings
for the nonlinear interaction in the gravitational field equations 
and the scalar field equation. 
$\alpha^{(s)}({\bf k}_1, {\bf k}_2)$, $\beta({\bf k}_1, {\bf k}_2)$ and $\gamma({\bf k}_1, {\bf k}_2)$ have the symmetry with respect to exchange between ${\bf k}_1$ and ${\bf k}_2$.
One can easily check that the functions to describe the nonlinear mode-couplings, 
$\alpha^{(s)}({\bf k}_1, {\bf k}_2), \beta({\bf k}_1, {\bf k}_2)$, and 
$\gamma({\bf k}_1, {\bf k}_2)$ satisfy
\begin{eqnarray}
\beta({\bf k}_1, {\bf k}_2) = \alpha^{(s)}({\bf k}_1, {\bf k}_2) - \gamma({\bf k}_1, {\bf k}_2).
\end{eqnarray}

\section{The third order equations}
In this section we consider the third order solutions. 
The third order solution of the cosmological density perturbations
has been investigated in various models~\cite{Juszkiewicz,Vishniac,Goroff,SutoSasaki,Makino,Bertshinger,
Scoccimarro,Takahashi,Komatsu,Koyama,fulltime}. 
We present the third order solution for the Horndeski's theory  in the cosmological background. 
Our results are general and applicable to various modified gravity models. 
Plus our results are useful for the case of the general relativity because 
we clarify the independence property of the mode-coupling functions and the relevant 
parameters to characterize the third order solution. 
We start with solving the third order equations for gravity and scalar field 
\begin{eqnarray}
&&-p^2\left({\cal F}_T\Psi_3(t,{\bf p})-{\cal G}_T\Phi_3(t,{\bf p})-A_1 Q_3(t,{\bf p})\right)
=\frac{B_1}{a^2H^2}\Gamma[t,{\bf p}; Q_1,Q_2]
+\frac{B_3}{a^2H^2}\Bigl(\Gamma[t,{\bf p}; Q_1,\Phi_2]
\nonumber\\
&&
\hspace{7cm}
+\Gamma[t,{\bf p}; Q_2,\Phi_1]\Bigr),
\label{thirdgra1}
\\
&&-p^2\left({\cal G}_T\Psi_3(t,{\bf p}) + A_2 Q_3(t,{\bf p})\right)
-\frac{a^2}{2}\rho_{\rm m}\delta_3(t,{\bf p})
=-\frac{B_2}{a^2H^2} 
\Gamma[t,{\bf p}; Q_1,Q_2]
-\frac{B_3}{a^2H^2}\Bigl(
\Gamma[t,{\bf p}; Q_1,\Psi_2] 
\nonumber\\
&&
\hspace{7cm}
+ \Gamma[t,{\bf p}; Q_2,\Psi_1]
\Bigr)
-\frac{C_1}{3a^4H^4}
\Xi_1[t,{\bf p};Q_1,Q_1,Q_1],
\label{thirdgra2}\\
&&-p^2(A_0Q_3(t,{\bf p})
-A_1\Psi_3(t,{\bf p})
-A_2\Phi_3(t,{\bf p}))
= - \frac{2 B_0}{a^2H^2}
\Gamma[t,{\bf p}; Q_1, Q_2]
+ \frac{B_1}{a^2H^2}\Bigl(
\Gamma[t,{\bf p}; Q_1, \Psi_2] 
\nonumber\\
&&
\hspace{1cm}
+ \Gamma[t,{\bf p}; Q_2,\Psi_1]\Bigr)
+ \frac{B_2}{a^2H^2}\Bigl(
\Gamma[t,{\bf p}; Q_1, \Phi_2] + \Gamma[t,{\bf p}; Q_2, \Phi_1]
\Bigr)
+ \frac{B_3}{a^2H^2}\Bigr(
\Gamma[t,{\bf p}; \Psi_1, \Phi_2] 
\nonumber\\
&&
\hspace{1cm}
+ \Gamma[t,{\bf p}; \Psi_2, \Phi_1]
\Bigr)
+ \frac{C_0}{a^4H^4}
\Xi_1[t,{\bf p};Q_1,Q_1,Q_1]
+ \frac{C_1}{a^4H^4}
\Xi_2[t,{\bf p};Q_1,Q_1,\Phi_1].
\label{thirdsca}
\end{eqnarray}
Inserting the first and the second order solutions into the above
equations, we finally have 
\begin{eqnarray}
&&{\cal F}_T\Psi_3(t,{\bf p})-{\cal G}_T\Phi_3(t,{\bf p})-A_1 Q_3(t,{\bf p})=- {a^2 H^2 \over p^2} D_+^3(t)
\Bigl(\left(B_1 \kappa_Q^2 + 2 B_3 \kappa_\Phi \kappa_Q \right){\cal W}_{\gamma\alpha}({\bf p}) 
\nonumber\\
&&~~~~~~~~~~~~~~~~
+\left(B_1 \kappa_Q \lambda_Q + B_3 \left(\kappa_\Phi\lambda_Q + \kappa_Q\lambda_\Phi\right)\right)
 {\cal W}_{\gamma\gamma}({\bf p})\Bigr),
\label{GRAA}
\\
&&{\cal G}_T\Psi_3(t,{\bf p}) + A_2 Q_3(t,{\bf p}) + {a^2 \over 2 p^2}\rho_{\rm m} \delta_3(t,{\bf p})
= {a^2 H^2 \over p^2} D_+^3(t)
\biggl(\left(B_2 \kappa_Q^2 + 2 B_3 \kappa_\Psi \kappa_Q \right){\cal W}_{\gamma\alpha}({\bf p}) 
\nonumber\\
&&~~~~~~~~~~~~~~~~+(B_2 \kappa_Q \lambda_Q + B_3 (\kappa_\Psi\lambda_{Q} + \kappa_Q \lambda_{\Psi 2}))
 {\cal W}_{\gamma\gamma}({\bf p})
 + {C_1\over 3} \kappa_Q^3{\cal W}_\xi({\bf p})\biggr),
\label{GRAB} 
\\
&& A_0Q_3(t,{\bf p})-A_1\Psi_3(t,{\bf p})-A_2\Phi_3(t,{\bf p})= - {a^2 H^2 \over p^2}D_+^3(t)
\biggl(\Bigl(-2B_0 \kappa_Q^2 + 2 B_1 \kappa_\Psi \kappa_Q 
\nonumber\\
&&~~~~~~~~~~~~~~~~
+ 2 B_2 \kappa_\Phi \kappa_Q + 2 B_3 \kappa_\Phi \kappa_\Psi\Bigr){\cal W}_{\gamma\alpha}({\bf p}) 
+ \Bigl((-2B_0 \kappa_Q \lambda_Q + B_1 (\kappa_\Psi \lambda_Q + \kappa_Q \lambda_\Psi)
\nonumber\\
&&~~~~~~~~~~~~~~~~
+ B_2 (\kappa_\Phi \lambda_Q + \kappa_Q \lambda_\Phi)
+ B_3 (\kappa_\Phi \lambda_\Psi + \kappa_\Psi\lambda_\Phi))
 {\cal W}_{\gamma\gamma}({\bf p})+(C_0 \kappa_Q^3 + C_1 \kappa_\Phi \kappa_Q^2){\cal W}_\xi({\bf p})\biggr),
\label{GRAC}
\end{eqnarray}
where we define ${\cal W}_{\gamma \alpha}({\bf p})$, ${\cal W}_{\gamma\gamma}({\bf p})$ and 
${\cal W}_\xi({\bf p})$ by Eqs.~(\ref{WGA}), (\ref{WGG}) and (\ref{WXI}), respectively,
in appendix B. 
Then, the gravitational and the curvature potentials, and the scalar field perturbations are written as
\begin{eqnarray}
\Phi_3(t,{\bf p})&=& - {a^2 H^2 \over p^2} \left(\kappa_\Phi(t) \delta_3(t,{\bf p}) + D_+^3(t)\left( \sigma_\Phi(t){\cal W}_{\gamma\alpha}({\bf p}) + \mu_\Phi(t) {\cal W}_{\gamma\gamma}({\bf p}) + \nu_\Phi(t) {\cal W}_{\xi}({\bf p})\right)\right),
\label{Phi3}
\\
\Psi_3(t,{\bf p})&=&- {a^2 H^2 \over p^2} \left(\kappa_\Psi(t) \delta_3(t,{\bf p}) + D_+^3(t)\left( \sigma_\Psi(t){\cal W}_{\gamma\alpha}({\bf p}) + \mu_\Psi(t) {\cal W}_{\gamma\gamma}({\bf p})+\nu_\Psi(t) {\cal W}_{\xi}({\bf p})\right)\right),
\\
Q_3(t,{\bf p})&=&- {a^2 H^2 \over p^2} \left(\kappa_Q(t) \delta_3(t,{\bf p}) + D_+^3(t)\left( \sigma_Q(t){\cal W}_{\gamma\alpha}({\bf p}) + \mu_Q(t) {\cal W}_{\gamma\gamma}({\bf p}) + \nu_Q(t) {\cal W}_{\xi}({\bf p})\right)\right),
\end{eqnarray}
where the coefficients $\sigma_\Phi(t)~\mu_\Phi(t),~\nu_\Phi(t)$, etc., 
are defined in appendix A. 
The third order equations for $\delta_3(t,{\bf p})$ and $\theta_3(t,{\bf p})$ are, 
\begin{eqnarray}
&&{1\over H}{\partial \delta_3(t,{\bf p})\over \partial t} + \theta_3(t,{\bf p}) \nonumber\\
&&\hspace{1.cm}= -{1\over (2\pi)^3}\int d{\bf k}_1 d{\bf k}_2 \delta^{(3)}({\bf k}_1 + {\bf k}_2 - {\bf p})\alpha({\bf k}_1,{\bf k}_2)(\theta_1(t,{\bf k}_1)\delta_2(t,{\bf k}_2) + \theta_2(t,{\bf k}_1)\delta_1(t,{\bf k}_2)),
\label{eoc1}\\
&&{1\over H}{\partial \theta_3(t,{\bf p})\over \partial t}+\left(2 + {\dot{H}\over H^2}\right)\theta_3(t,{\bf p})-{p^2\over a^2 H^2} \Phi_3(t,{\bf p})\nonumber\\
&&\hspace{1.cm}= -{2\over (2\pi)^3}\int d{\bf k}_1 d{\bf k}_2 \delta^{(3)} ({\bf k}_1+ {\bf k}_2-{\bf p})\beta({\bf k}_1,{\bf k}_2)\theta_1(t,{\bf k}_1)\theta_2(t,{\bf k}_2).\label{eoe1}
\end{eqnarray}
Using the first and the second order solutions, these equations are rewritten as
\begin{eqnarray}
&&{1\over H}{\partial \delta_3(t,{\bf p})\over \partial t} + \theta_3(t,{\bf p}) \nonumber\\
&&\hspace{1.5cm}= D_+^3(t) f \left({\cal W}_{\alpha\alpha R}({\bf p}) - {2\over 7} \lambda {\cal W}_{\alpha\gamma R}({\bf p}) + {\cal W}_{\alpha\alpha L}({\bf p})- {4\over 7} \lambda_\theta  {\cal W}_{\alpha\gamma L}({\bf p}) \right),
\label{eoc33}\\
&&{1\over H}{\partial \theta_3(t,{\bf p})\over \partial t}+\left(2 + {\dot{H}\over H^2}\right)\theta_3(t,{\bf p})-{p^2\over a^2 H^2} \Phi_3(t,{\bf p})\nonumber\\
&&\hspace{1.5cm}= 2 D_+^3(t) f^2 \left(- {\cal W}_{\alpha\alpha}({\bf p}) + {4\over 7} \lambda_\theta {\cal W}_{\alpha\gamma}({\bf p}) +  {\cal W}_{\gamma\alpha}({\bf p}) - {4\over 7} \lambda_\theta {\cal W}_{\gamma\gamma}({\bf p}) \right),
\label{eoe33}
\end{eqnarray}
where we introduce the functions defined by Eqs. (\ref{WAAR}) to (\ref{WAR}),  
for which we find that the following relations hold,
\begin{eqnarray}
2{\cal W}_{\gamma\alpha}({\bf p}) &=& {\cal W}_{\alpha\gamma R}({\bf p}) + 2{\cal W}_{\gamma\gamma}({\bf p}) - {\cal W}_\xi({\bf p}),\label{rel1}\\
2{\cal W}_{\alpha\alpha}({\bf p}) &=& {\cal W}_{\alpha\alpha R}({\bf p}) + {\cal W}_{\alpha\alpha L}({\bf p}),\label{rel2}\\
2{\cal W}_{\alpha\gamma}({\bf p}) &=& {\cal W}_{\alpha\gamma R}({\bf p}) + {\cal W}_{\alpha\gamma L}({\bf p}).\label{rel3}
\end{eqnarray}
Then, Eqs.~(\ref{eoc33}) and (\ref{eoe33}) reduce to
\begin{eqnarray}
&&{1\over H}{\partial \delta_3(t,{\bf p})\over \partial t} + \theta_3(t,{\bf p}) 
= D_+^3(t) f \left(2 {\cal W}_{\alpha\alpha}({\bf p}) - {2\over 7} \lambda {\cal W}_{\alpha\gamma R}({\bf p}) - {4\over 7} \lambda_\theta  {\cal W}_{\alpha\gamma L}({\bf p}) \right),
\label{3rdconeq}\\
&&{1\over H}{\partial \theta_3(t,{\bf p})\over \partial t}+\left(2 + {\dot{H}\over H^2}\right)\theta_3(t,{\bf p})-{p^2\over a^2 H^2} \Phi_3(t,{\bf p})
= 2 D_+^3(t) f^2 \biggl(- {\cal W}_{\alpha\alpha}({\bf p}) 
\nonumber\\
&&~~~~~~~~~~~~~~~~~~
+ \Bigl({2\over 7} \lambda_\theta + {1\over 2}\Bigr) {\cal W}_{\alpha\gamma R}({\bf p}) +
{2\over 7} \lambda_\theta {\cal W}_{\alpha\gamma L}({\bf p}) + \Bigl(1 - {4\over 7} \lambda_\theta \Bigr) 
{\cal W}_{\gamma\gamma}({\bf p}) - {1\over 2} {\cal W}_\xi({\bf p})\biggr).
\label{eoe33b}
\end{eqnarray}
Combining these two equations, we have the third order equation for $\delta_3(t, {\bf p})$ as
\begin{eqnarray}
{\partial^2\delta_3(t,{\bf p}) \over \partial t^2} + 2 H {\partial \delta_3(t,{\bf p}) \over \partial t} + L(t) \delta_3(t,{\bf p}) = S_{\delta3}(t,{\bf p}),
\label{3rddeq}
\end{eqnarray}
where we define
\begin{eqnarray}
S_{\delta3}(t) &=& D_+^3(t)\Bigl(N_{\alpha\alpha}(t){\cal W}_{\alpha\alpha}({\bf p}) + N_{\alpha\gamma R}(t){\cal W}_{\alpha\gamma R}({\bf p}) + N_{\alpha\gamma L}(t){\cal W}_{\alpha\gamma L}({\bf p})
\nonumber\\
&&~~+ N_{\gamma\gamma}(t){\cal W}_{\gamma\gamma}({\bf p}) + N_\xi(t){\cal W}_\xi({\bf p}) \Bigr),
\label{3rdnh}
\end{eqnarray}
and 
\begin{eqnarray}
&&N_{\alpha\alpha}(t) = 6 f^2 H^2 - 2 L\label{nalal},\\
&&N_{\alpha\gamma R}(t) = -f^2 H^2 - {8\over 7}f^2 H^2 \lambda + {2\over 7}L \lambda - {4\over 7}f H \dot{\lambda}+{1\over 2}H^2 \sigma_\Phi\label{nalgaR},\\
&&N_{\alpha\gamma L}(t) 
=- f^2 H^2 - {8\over 7}f^2 H^2\lambda + {2\over 7}L \lambda - {4\over 7}f 
H \dot{\lambda} + N_\gamma
\label{NAGL},\\
&&N_{\gamma\gamma}(t) = -2 f^2 H^2 + {8\over 7} f^2 H^2 \lambda + {4\over 7} f H \dot{\lambda}  + H^2 \left( \sigma_\Phi +\mu_\Phi \right),\label{ngaga}\\
&&N_\xi(t) =  f^2 H^2 + H^2 \left(- {1\over 2} \sigma_\Phi + \nu_\Phi \right). \label{nxi}
\end{eqnarray}
where used Eqs.~(\ref{lambdatheta}) and (\ref{2ndlambdade}), and 
\begin{eqnarray}
\dot{f}(t) &=& {1\over H}(- 2 f H^2 - L - f^2 H^2 - f \dot{H}),
\end{eqnarray}
which follow from the definition of the growth rate 
$f(t)=d\ln D_+/d\ln a$ and Eq.~(\ref{linearDP}). 
We can prove that $N_{\alpha\gamma L}(t)$ is equivalent to $N_{\alpha\gamma R}(t)$, 
using (\ref{NAGL}) and (\ref{nalgaR}), and $N_\gamma(t) = {1\over 2}H^2 \sigma_\Phi$,
which is demonstrated from Eqs.~(\ref{NGAMMA}) and (\ref{SIGMAPHI}). 
Then, we write
\begin{eqnarray}
N_{\alpha\gamma}(t) \equiv 
N_{\alpha\gamma R}(t) = N_{\alpha\gamma L}(t).
\label{reql}
\end{eqnarray}

The general solution of Eq.~(\ref{3rddeq}) with (\ref{3rdnh}) is 
\begin{eqnarray}
\delta_3(t,{\bf p}) &=& c_+({\bf p}) D_+(t) + c_-({\bf p})D_-(t) + \int^t_0 {D_-(t)D_+(t') - D_+(t)D_-(t')\over W(t')} S_{\delta3}(t', {\bf p})dt',
\end{eqnarray}
where $D_+(t)$ and $D_-(t)$ are the growing mode solution and the 
decaying mode solution, satisfying equation (\ref{linearDP}), 
$c_+({\bf p})$ and $c_-({\bf p})$ are integral constants, and
$W(t)$ is the Wronskian defined by $W(t)= D_+(t) \dot{D}_-(t) - \dot{D}_+(t) D_-(t)$. 
Since we assume that the initial density perturbations obey the Gauss distribution, 
we set $c_\pm({\bf p}) = 0$, as is done in deriving the second order solution. 
Then, the solution of the third order density perturbations is given by 
\begin{eqnarray}
&&\delta_3(t, {\bf p}) = D_+^3(t)\biggl(\kappa_{\delta3}(t){\cal W}_{\alpha\alpha}({\bf p}) - {2\over 7}\lambda_{\delta3}(t){\cal W}_{\alpha\gamma R}({\bf p}) - {2\over 7}\lambda_{\delta3}(t){\cal W}_{\alpha\gamma L}({\bf p}) 
\nonumber\\
&&~~~~~~~~~~~~~~~
- {2\over 21} \mu(t){\cal W}_{\gamma\gamma}({\bf p}) + {1\over 9} \nu(t) {\cal W}_\xi({\bf p})\biggr),
\label{3rddel}
\end{eqnarray}
where we define
\begin{eqnarray}
\kappa_{\delta3}(t) &=&  {1\over D_+^3(t)}\int^t_0 {D_-(t)D_+(t') - D_+(t)D_-(t')\over W(t')}D_+^3(t') N_{\alpha\alpha}(t')dt' \label{3rdkappa},\\
\lambda_{\delta3}(t) &=&  - {7\over 2 D_+^3(t)}\int^t_0 {D_-(t)D_+(t') - D_+(t)D_-(t')\over W(t')}D_+^3(t') N_{\alpha\gamma}(t')dt', \label{3rdlambda}\\
\mu(t) &=&  - {21\over 2 D_+^3(t)}\int^t_0 {D_-(t)D_+(t') - D_+(t)D_-(t')\over W(t')}D_+^3(t') N_{\gamma\gamma}(t')dt', \label{3rdmu} \\
\nu(t) &=&  {9\over  D_+^3(t)} \int^t_0 {D_-(t)D_+(t') - D_+(t)D_-(t')\over W(t')}D_+^3(t') N_\xi(t')dt'. \label{3rdnu}
\end{eqnarray}
Here note that the parameters in front of ${\cal W}_{\alpha\gamma R}({\bf p})$ and 
${\cal W}_{\alpha\gamma L}({\bf p})$ in expression (\ref{3rddel}) are the same, 
which originates from the relation~(\ref{reql}). 
In the limit of the Einstein de Sitter universe in the general relativity, 
the coefficients, $\kappa_{\delta3}(t)$, $\lambda_{\delta3}(t)$, $\mu(t)$, and $\nu(t)$
reduce to $1$.

We can redefine these coefficients using the differential equations. 
Inserting the general form of the solution (\ref{3rddel}) into (\ref{3rddeq}), 
we obtain the following differential equations for the coefficients
\begin{eqnarray}
&&\ddot{\kappa}_{\delta3}(t) + (6 f + 2)\dot{\kappa}_{\delta3}(t) + (6 f^2 H^2 - 2 L)\kappa_{\delta3}(t) = N_{\alpha\alpha}(t),
\label{DEA}\\
&&\ddot{\lambda}_{\delta3}(t) + (6 f + 2)\dot{\lambda}_{\delta3}(t) + (6 f^2 H^2 - 2 L)\lambda_{\delta3}(t) = -{7\over 2}N_{\alpha\gamma}(t),
\label{DEB}\\
&&\ddot{\mu}(t) + (6 f + 2)\dot{\mu}(t) + (6 f^2 H^2 - 2 L)\mu(t) = -{21\over 2}N_{\gamma\alpha}(t),\\
&&\ddot{\nu}(t) + (6 f + 2)\dot{\nu}(t) + (6 f^2 H^2 - 2 L)\nu(t) = 9 N_\xi(t).
\label{DED}
\end{eqnarray}
The homogeneous solution of all these equations is $1/D_+^2(t)$ and $D_-(t)/D_+^3(t)$.
Therefore, the differential equations (\ref{DEA}) to (\ref{DED}) consistently yield 
the inhomogeneous solutions (\ref{3rdkappa}) to (\ref{3rdnu}), respectively. 

We next show that $\kappa_{\delta3}(t)=1$ identically. 
Using the expression (\ref{nalal}), we easily find that $\kappa_{\delta3}=1$ is the 
solution of (\ref{DEA}). This means that the inhomogeneous solution (\ref{3rdkappa})
reduces to $\kappa_{\delta3}=1$. We can prove $\kappa_{\delta3} = 1$ directly 
from (\ref{3rdkappa}), using partial integral.

Furthermore we can show that $\lambda_{\delta3}(t)=\lambda(t)$ identically. 
We can rewrite Eq.~(\ref{DEB}), as follows,
\begin{eqnarray}
&&\ddot{\lambda}_{\delta3}(t) + (4 f + 2)H\dot{\lambda}_{\delta3}(t) + (2 f^2 H^2 -  L)\lambda_{\delta3}(t) + 2 f H (\dot{\lambda}_{\delta3} - \dot{\lambda}) 
\nonumber\\
&&~~~~~~~~~~~~~~~~~~~~~~~~~~~~~~
+ (4 f^2 H^2 - L) (\lambda_{\delta3} - \lambda) = {7\over 2}(f^2 H^2 - N_\gamma),
\end{eqnarray}
where we used (\ref{reql}) and (\ref{NAGL}).
We can easily check that $\lambda_{\delta3}(t)$ and $\lambda(t)$ 
satisfies the same differential equation (see Eq.~(\ref{2ndlambdade})), which leads to 
$\lambda_{\delta3}(t)=\lambda(t)$. 

In summary, we have the expression equivalent to (\ref{3rddel}),
\begin{eqnarray}
\delta_{3}(t, {\bf p}) = D_+^3(t)\left({\cal W}_{\alpha\alpha}({\bf p}) - {2\over7}\lambda(t){\cal W}_{\alpha\gamma R}({\bf p}) - {2\over7}\lambda(t){\cal W}_{\alpha\gamma L}({\bf p}) - {2\over 21}\mu(t){\cal W}_{\gamma\gamma}({\bf p}) + {1\over 9}\nu(t){\cal W}_\xi({\bf p})\right).
\label{eoc33f}
\end{eqnarray}
Thus the third order solution of density contrast is 
characterized by $\lambda(t)$, $\mu(t)$, and $\nu(t)$. 
Note that $\lambda(t)$ is defined to describe the second order solution, then $\mu(t)$ and $\nu(t)$ are 
the new coefficients which appear at the third order. 
Table I summarizes the parameters and the mode-coupling functions necessary to 
describe the second order solution and the third order solution. 

Recently, the authors of \cite{fulltime} investigated the third order solution of 
the density perturbations, in a similar way, but within a model of the
general relativity. In their paper, 6 parameters are introduced to describe the 
third order density perturbations. Our results suggest that less number of 
parameters are only independent.

Inserting the solution (\ref{eoc33f}) into Eq.~(\ref{3rdconeq}), we find the solution for  the 
velocity divergence
\begin{eqnarray}
\theta_3(t, {\bf p}) &=& - D_+^3(t) f \biggl({\cal W}_{\alpha\alpha}({\bf p}) - {4\over 7}\lambda_\theta(t){\cal W}_{\alpha\gamma R} - {2\over 7}\lambda (t){\cal W}_{\alpha\gamma L}({\bf p}) 
\nonumber\\
&&~~~~~~~~~~~~~~~
- {2\over 7}\mu_\theta(t){\cal W}_{\gamma\gamma}({\bf p}) + {1\over 3}\nu_\theta(t){\cal W}_\xi({\bf p})\biggr),
\label{eoe33f}
\end{eqnarray}
where we define
\begin{eqnarray}
\mu_{\theta}(t) &=& \mu(t) + {\dot{\mu}(t)\over 3 f H}, \label{mutheta}\\
\nu_{\theta}(t) &=& \nu(t) + {\dot{\nu}(t)\over 3 f H}. \label{nutheta}
\end{eqnarray}
Here note that $\lambda_\theta(t)$ is the 
parameter to describe the second order solution, and 
$\mu_\theta(t)$ and $\nu_\theta(t)$ are the new parameters which appear at the third order.  \\

In summary, we first introduced {\it nine} mode-coupling functions in the third order 
equations, 
(\ref{eoc33}) and (\ref{eoe33}) with (\ref{Phi3}). 
We find the {\it three} identities (\ref{rel1}), (\ref{rel2}) and (\ref{rel3}).
Then, only {\it six} mode-coupling functions are independent in the {\it nine} ones. 
This conclusion that the number of the linearly independent mode-coupling functions 
is {\it six} can be proved by using the generalized Wronskian. 
The coefficients in front of ${\cal W}_{\alpha\alpha R}$ and ${\cal W}_{\alpha\alpha L}$
in equation (\ref{eoc33}) are the same, which leads to the final third order solution
(\ref{eoc33f}) and (\ref{eoe33f}) expressed in terms of the {\it five} mode-coupling functions. 

\begin{table}[t] 
\begin{tabular}{r|l|c} \hline\hline
 ~~~~~~~~~~~~& ~~~~~~~~~~parameters~~~~~~~~~ & ~~~~~~~mode-coupling functions~~~~~~~ 
\\ \hline$\delta_2(t, {\bf p})$ & $\lambda(t)$ & ${\cal W}_\alpha({\bf p})$ 
\\ \cline{1-1} \cline{2-2}
$\theta_2(t, {\bf p})$ & $\lambda_\theta(t)$ & ${\cal W}_\gamma({\bf p})$ 
\\ \hline
$\delta_3(t, {\bf p})$ & $\lambda(t), ~{\mu(t), ~\nu(t)}$ & ${\cal W}_{\alpha\alpha}({\bf p}), ~{\cal W}_{\alpha\gamma R}({\bf p}), ~{\cal W}_{\alpha\gamma L}({\bf p})$ 
\\ \cline{1-1} \cline{2-2}
$\theta_3(t, {\bf p})$ & $\lambda(t), ~\lambda_\theta(t),  ~{\mu_\theta(t), ~\nu_\theta(t)}$
& ${\cal W}_{\gamma\gamma}({\bf p}), ~{\cal W}_\xi({\bf p})$ 
 \\ \hline\hline
\end{tabular}
\caption{Functions for the mode-couplings and parameters necessary to describe
the second order solution and the third order solution.}
\end{table}

\section{Power spectrum}
The third order solution of the density perturbations enable one to compute
the 1-loop (second order) power spectrum. The second order matter power 
spectrum was computed by many authors \cite{Juszkiewicz,Vishniac,Goroff,SutoSasaki,Makino,Bertshinger,
Scoccimarro,Takahashi,Komatsu,Koyama,fulltime}, in general relativity and modified gravity models. 
We find the expression for the 1-loop order the power spectra of density contrast and 
velocity divergence by
\begin{eqnarray}
\left<\delta(t,{\bf k}_1)\delta(t,{\bf k}_2)\right> &=&(2 \pi)^3\delta^{(3)}({\bf k}_1 + {\bf k}_2)P_{\delta\delta}(t, k),
\label{definedd}
\\
\left<\delta(t,{\bf k}_1)\theta(t,{\bf k}_2)\right> &=&(2 \pi)^3\delta^{(3)}({\bf k}_1 + {\bf k}_2) (- f) P_{\delta\theta}(t,k),
\label{definedt}
\\
\left<\theta(t,{\bf k}_1)\theta(t,{\bf k}_2)\right> &=&(2 \pi)^3\delta^{(3)}({\bf k}_1 + {\bf k}_2) f^2 P_{\theta\theta}(t, k),
\label{definett}
\end{eqnarray}
where we use the notation $k = |{\bf k}_1|$. 
Some details of their derivations are described in appendix C, we here show the
results,
\begin{eqnarray}
P_{\delta\delta}(t, k) &=& D_+^2(t) P_{\rm L}(k) + D_+^4(t)\left(P_{\delta\delta}^{(22)}(t, k) + 2 P_{\delta\delta}^{(13)}(t, k)\right),\\
P_{\delta\theta}(t, k) &=& D_+^2(t) P_{\rm L}(k) + D_+^4(t)\left(P_{\delta\theta}^{(22)}(t, k) + 2 P_{\delta\theta}^{(13)}(t, k)\right),\\
P_{\theta\theta}(t, k) &=& D_+^2(t) P_{\rm L}(k) + D_+^4(t)\left(P_{\theta\theta}^{(22)}(t, k) + 2 P_{\theta\theta}^{(13)}(t, k)\right),
\end{eqnarray}
where $D^2_+(t)P_{\rm L}(k)$ is the linear matter power spectrum,  and we define 
\begin{eqnarray}
&&P_{\delta\delta}^{(22)}(t,k) = {k^3 \over 98(2\pi)^2}\int d r P_{\rm L}(rk)\int^1_{-1} dx P_{\rm L}(k(1 + r^2 - 2 rx)^{1/2})
\nonumber\\
&&~~~~~~~~~~~~\times
{((7- 4 \lambda) r + 7 x + 2(2\lambda - 7)rx^2)^2\over (1 + r^2 - 2  r x)^2},
\label{Pdd}
\\
&&P_{\delta\theta}^{(22)}(t, k)= {k^3 \over 98(2\pi)^2}\int d r P_{\rm L}(rk)\int^1_{-1} dx P_{\rm L}(k(1 + r^2 - 2  r x)^{1/2})
\nonumber\\
&&~~~~~~~~~~~~\times
{((7- 4 \lambda) r + 7  x + 2(2\lambda - 7)r x^2)((7- 8 \lambda_\theta) r + 7  x + 2(4\lambda_\theta - 7) r x^2)\over (1 + r^2 - 2rx)^2}
\label{Pdt}
\\
&&P_{\theta\theta}^{(22)}(t, k) = {k^3 \over 98(2\pi)^2}\int d r P_{\rm L}(rk)\int^1_{-1} dx P_{\rm L}(k(1 + r^2 - 2 r x)^{1/2})
\nonumber\\
&&~~~~~~~~~~~~\times
{((7- 8 \lambda_\theta) r + 7  x + 2(4\lambda_\theta - 7) r x^2)^2\over (1 + r^2 - 2 r x)^2},
\label{Ptt}
\end{eqnarray}
and
\begin{eqnarray}
&&2P_{\delta\delta}^{(13)}(t,k) = { k^3 \over252(2\pi)^2 }P_{\rm L}(k)\int d r P_{\rm L}(rk) 
\biggl[12 \mu {1\over r^2} - 2 (21 + 36 \lambda + 22\mu) 
+ 4(84 - 48 \lambda - 11\mu){r^2} 
\nonumber\\
&&~~~~~~~~~~~~~~~~~~ - 6(21 - 12 \lambda - 2 \mu) {r^4}+ {3\over r^3} \left(r^2 - 1\right)^3 \left((21 - 12 \lambda - 2 \mu)r^2 + 2 \mu \right)\ln \left( {r + 1\over |r - 1|}\right) \biggr], \label{P13MGST}
\\
&&2P_{\delta\theta}^{(13)}(t,k) = { k^3 \over 252(2\pi)^2}P_{\rm L}(k)\int d r P_{\rm L}(rk) 
\biggl[6 (\mu + 3\mu_\theta) {1\over r^2} - 2 (21 + 36 \lambda + 11\mu + 33\mu_\theta)  
\nonumber\\
&& \hspace{2cm} + 2(168 - 96 \lambda - 11\mu - 33\mu_\theta){r^2} - 6(21 - 12 \lambda - \mu - 3\mu_\theta) {r^4} 
\nonumber\\
&& \hspace{2cm}+ {3\over r^3} \left(r^2 - 1\right)^3 \left((21 - 12 \lambda - \mu - 3 \mu_\theta)r^2 + (\mu + 3\mu_\theta) \right)\ln \left( {r + 1\over |r - 1|}\right) \biggr],
\label{P13dt}
\\
&&2P_{\theta\theta}^{(13)}(t,k) = {k^3 \over 84(2\pi)^2}P_{\rm L}(k)\int d r P_{\rm L}(rk) 
\biggl[12 \mu_\theta {1\over r^2} - 2 (7 + 12 \lambda + 22\mu_\theta) + 4(28 - 16 \lambda - 11\mu_\theta){r^2} 
\nonumber\\
&& \hspace{2cm}
- 6(7 - 4 \lambda - 2 \mu_\theta) {r^4} 
+ {3\over r^3} \left(r^2 - 1\right)^3 \left((7 - 4 \lambda - 2 \mu_\theta)r^2 + 2 \mu_\theta \right)\ln \left( {r + 1\over |r - 1|}\right) \biggr].
\label{P13tt}
\end{eqnarray}
The third order solutions of the density contrast and the velocity divergence are 
described by 6 parameters in Table I. 
The 1-loop power spectra are described by 4 parameters, and they do not depend on 
$\nu(t)$ and $\nu_\theta(t)$ (see Table II). 
In deriving the 1-loop power spectrum, we find that the relation,
\begin{eqnarray}
\xi({\bf k}, {\bf q}_1, -{\bf q}_1) = 0,
\end{eqnarray}
holds, which prevents the 1-loop power spectrum from depending on $\nu(t)$ and $\nu_\theta(t)$.
Details are described in appendix C and D.

\begin{table}[b] 
\begin{tabular}{c|l} \hline\hline
~~~~~~~  & ~~~parameters~~~~~~~~ 
\\ \hline
$P_{\delta\delta}$ & $\lambda(t), ~\mu(t)$
\\ \hline
$P_{\delta\theta}$ & $\lambda(t), ~\mu(t), ~\lambda_\theta(t), ~\mu_\theta(t)$
\\ \hline
$P_{\theta\theta}$ & $\lambda(t), ~\lambda_\theta(t), ~\mu_\theta(t)$
\\ \hline\hline
\end{tabular}
\caption{Summary of the parameters to characterize the 1-loop order power spectra
$P_{\delta\delta}$, $P_{\delta\theta}$, and $P_{\theta\theta}$, respectively.}
\end{table}

\section{Trispectrum}
Here we present the expression for the matter
trispectrum in the real space, which is defined by  
\begin{eqnarray}
\left<\delta(t,{\bf k}_1)\delta(t,{\bf k}_2)\delta(t,{\bf k}_3)\delta(t,{\bf k}_4)\right>_c =
(2 \pi)^3\delta^{(3)}({\bf k}_1 + {\bf k}_2 + {\bf k}_3 + {\bf k}_4)T(t,{\bf k}_1,{\bf k}_2,{\bf k}_3,{\bf k}_4).
\end{eqnarray}
Using the solution up to the third order of the density perturbations, we find
\begin{eqnarray}
T(t,{\bf k}_1,{\bf k}_2,{\bf k}_3,{\bf k}_4)
=D_+^6(t)\Bigl(T_{1122}({\bf k}_1,{\bf k}_2,{\bf k}_3,{\bf k}_4)
+T_{1113}({\bf k}_1,{\bf k}_2,{\bf k}_3,{\bf k}_4)\Bigr),
\end{eqnarray}
where we define
\begin{eqnarray}
&&T_{1122}(t,{\bf k}_1,{\bf k}_2,{\bf k}_3,{\bf k}_4) = 4 P_{11}(k_1)P_{11}(k_2) \left[P_{11}(|{\bf k}_1 + {\bf k}_3|) F_2(t,{\bf k}_1,- {\bf k}_1 - {\bf k}_3)F_2(t,{\bf k}_2,{\bf k}_1 + {\bf k}_3) 
\right.\nonumber\\
&&\hspace{3.5cm}\left. + P_{11}(|{\bf k}_1 + {\bf k}_4|)F_2(t,{\bf k}_1, - {\bf k}_1 - {\bf k}_4)F_2(t,{\bf k}_2,{\bf k}_1 + {\bf k}_4)\right] + 5~{\rm cyclic~terms},
\\
&&T_{1113}(t,{\bf k}_1,{\bf k}_2,{\bf k}_3,{\bf k}_4)=6P_{11}(k_1)P_{11}(k_2)P_{11}(k_3) F_3(t,{\bf k}_1,{\bf k}_2, {\bf k}_3) + 4 ~{\rm cyclic~terms}. 
\end{eqnarray}
Note that $T_{1122}$ contains $F_{2}$, defined by Eq.~(\ref{efu2}), which depends on $\lambda(t)$, 
while $T_{1113}$ contains $F_3$, defined by Eq.~(\ref{efu3}), which depends on $\lambda(t)$, 
$\mu(t)$ and $\nu(t)$. Therefore, the matter trispectrum depends on these three parameters.
\section{Application of KGB model}
In this section, we exemplify the effect of the modified gravity on 
the 1-loop power spectrum. We here consider the kinetic 
gravity braiding (KGB) model \cite{Deffayet,kgb}, which
is considered in Ref.~\cite{TTY} to demonstrate the effect of the 
modified gravity on the bispectrum. We briefly review the model. 
The action of the KGB model is written as
\begin{eqnarray}
S=\int d^4 x\sqrt{-g}\left[{M_{\rm pl}^2\over 2}R+K(\phi, X)
-G_3(\phi, X)\square \phi+{\cal L}_{\rm m}\right], 
\label{kgbaction}\end{eqnarray}
where $M_{\rm pl}$ is the Planck mass, which is related with the gravitational constant $G_N$
by $8\pi G_N=1/M_{\rm pl}^2$. Comparing this action (\ref{kgbaction}) with that of the 
most general second-order scalar-tensor theory (\ref{action}), the action of the kinetic gravity 
braiding model is produced by setting 
\begin{eqnarray}
&&G_4={M_{\rm pl}^2\over 2},~~~~G_5=0,
\end{eqnarray}
and we choose $K$ and $G_3$ as
\begin{eqnarray}
&&K=-X, ~~~~G_3=M_{\rm pl}\left({r_c^2\over M_{\rm pl}^2}X\right)^n,  
\end{eqnarray}
where $n$ and $r_c$ are the model parameters. 
Useful expressions of the kinetic gravity braiding model are summarized in appendix E.

When we consider the attractor  solution, 
which satisfies $3\dot\phi H G_{3X}=1$,  
the Friedmann equation is written in the form 
\begin{eqnarray}
\left({H\over H_0}\right)^2={\Omega_0\over a^3}
+(1-\Omega_0)\left({H\over H_0}\right)^{-2/(2n-1)},
\label{habbulu}
\end{eqnarray}
where $H_0$ is the Hubble constant and 
$\Omega_0$ is the density parameter
at the present time, and the model parameters must satisfy
\begin{eqnarray}
H_0r_c=\left({2^{n-1}\over 3n}\right)^{1/2n}\left[{1\over 6(1-\Omega_0)}\right]^{(2n-1)/4n}. 
\end{eqnarray}
On the attractor solution, $L(t)$, defined by Eq.~(\ref{elut}), reduces to
\begin{eqnarray}
&&L(t)=-{3\over 2}{2n + (3n -1)\Omega_{\rm m}\over 5n-\Omega_{\rm m}}H^2,
\label{kgbL}
\end{eqnarray}
where $\Omega_{\rm m}$ is defined by $\Omega_{\rm m}(a)={\Omega_0H_0^2/H(a)^2a^3}$. 
The linear growth factor $D_+$ is obtained from Eq.~(\ref{linearDP}) with (\ref{kgbL}) 
and (\ref{habbulu}).
However, note that the quasi-static approximation on the scales of the large scale 
structures  holds for $n\simlt 10$ (see \cite{kgb}). 

The second order solution and the third order solution are obtained with
\begin{eqnarray}
&&N_{\gamma}(t)={1\over 2} H^2  \sigma_\Phi(t) = -{9\over 4}{(1 - \Omega_{\rm m})(2n - \Omega_{\rm m})^3\over \Omega_{\rm m}(5n - 
\Omega_{\rm m})^3}H^2,\label{kgbngamma}\\
&&H^2 \mu_\Phi(t) ={9  (1 - \Omega_{\rm m}) (2 n - \Omega_{\rm m})^3 (4 n^2 (21 + 25 \lambda \Omega_{\rm m}) - 4 n \Omega_{\rm m} (21 + 10 \lambda \Omega_{\rm m})+\Omega_{\rm m}^2 (21 + 4 \lambda \Omega_{\rm m})) \over 28 \Omega_{\rm m}^2 (5 n - \Omega_{\rm m})^5}H^2, \nonumber\\
\\
&&\nu_\Phi(t) = 0.
\end{eqnarray}
We have $\lambda(t)$ from (\ref{2ndlambda}) with (\ref{kgbngamma}). 
Using these results and Eqs.~(\ref{ngaga}) and (\ref{nxi}), 
we have the expressions for $\mu(t)$ and $\nu(t)$ from (\ref{3rdmu}) and (\ref{3rdnu}). 
Eqs.~(\ref{lambdatheta}), (\ref{mutheta}), and (\ref{nutheta}) give expressions 
for $\lambda_\theta(t), ~\mu_\theta(t), ~\nu_\theta(t)$, respectively. 

Table III lists the numerical values of these variables at the redshift $z=1$, $0.5$ 
and $0$, for the KGB model with $n=1$, $2$, $5$, as well as the $\Lambda $CDM model. 

\begin{table}[b]
  \begin{tabular}{r|c|c|c|c} \hline\hline
   ~~~~~~~~~~~~~~~~~~~~~~~ & ~~~~~~~~~ $ \Lambda$CDM ~~~~~~~~~ & ~~~~~ ${\rm KGB}(n = 1)$ ~~~~~ & ~~~~~ ${\rm KGB}(n = 2)$ ~~~~~ &~~~~~~  ${\rm KGB}(n = 5)$ ~~~~~  \\ \hline 
    $D_+(z = 1~/~0.5~/~0)$ &       $0.477~/~0.602~/~0.779$ & $0.496~/~0.642~/~0.858$ & $0.489~/~0.628~/~0.838$ & $0.484~/~0.620~/~0.827$ \\ \hline
    $f(z = 1~/~0.5~/~0)$ &         $0.869~/~0.749~/~0.513$ & $0.951~/~0.835~/~0.593$ & $0.919~/~0.813~/~0.605$ & $0.904~/~0.805~/~0.612$ \\ \hline
    $\lambda(z = 1~/~0.5~/~0)$ &   $0.999~/~0.997~/~0.994$ & $1.000~/~0.999~/~1.003$ & $1.000~/~1.000~/~1.011$ & $1.000~/~1.002~/~1.019$ \\ \hline
    $\mu(z = 1~/~0.5~/~0)$ &       $0.999~/~0.998~/~0.996$ & $1.000~/~1.001~/~1.015$ & $1.001~/~1.005~/~1.015$ & $1.003~/~1.007~/~1.011$ \\ \hline
    $\nu(z = 1~/~0.5~/~0)$ &       $0.998~/~0.996~/~0.991$ & $1.000~/~0.999~/~1.014$ & $1.000~/~1.003~/~1.034$ & $1.002~/~1.008~/~1.049$ \\ \hline
    $\lambda_\theta(z=1~/~0.5~/~0)$&$0.994~/~0.991~/~0.983$ & $0.998~/~0.995~/~1.043$ & $0.999~/~1.004~/~1.073$ & $1.003~/~1.014~/~1.095$ \\ \hline 
    $\mu_\theta(z = 1~/~0.5~/~0)$ & $0.997~/~0.995~/~0.991$ & $1.000~/~1.006~/~1.041$ & $1.006~/~1.018~/~1.008$ & $1.010~/~1.021~/~0.974$ \\ \hline 
    $\nu_\theta(z = 1~/~0.5~/~0)$ & $0.994~/~0.990~/~0.980$ & $0.998~/~0.998~/~1.089$ & $1.002~/~1.014~/~1.136$ & $1.009~/~1.030~/~1.169$ \\ \hline\hline
  \end{tabular}
    \caption{Numerical values of the growth factor $D_+$, the linear growth rate $f$, and the coefficients $\lambda, ~\mu, ~\nu$, ~$\lambda_\theta, ~\mu_\theta, ~\nu_\theta$ at the redshifts $z=1.0$, ~$0.5$ and $0$, for 
the $\Lambda \rm CDM$ mode and the ${\rm KGB}$ model with $n = 1, ~2, ~5$. 
In each cell, a set of the three numerics means the 
values at the redshift $z=1.0$, ~$0.5$ and $0$ from left to right, respectively.}
\end{table}

\begin{figure}[h!]
  \begin{center}
   \vspace{5.0cm}
  \includegraphics[width=180mm,bb=0 0 640 480]{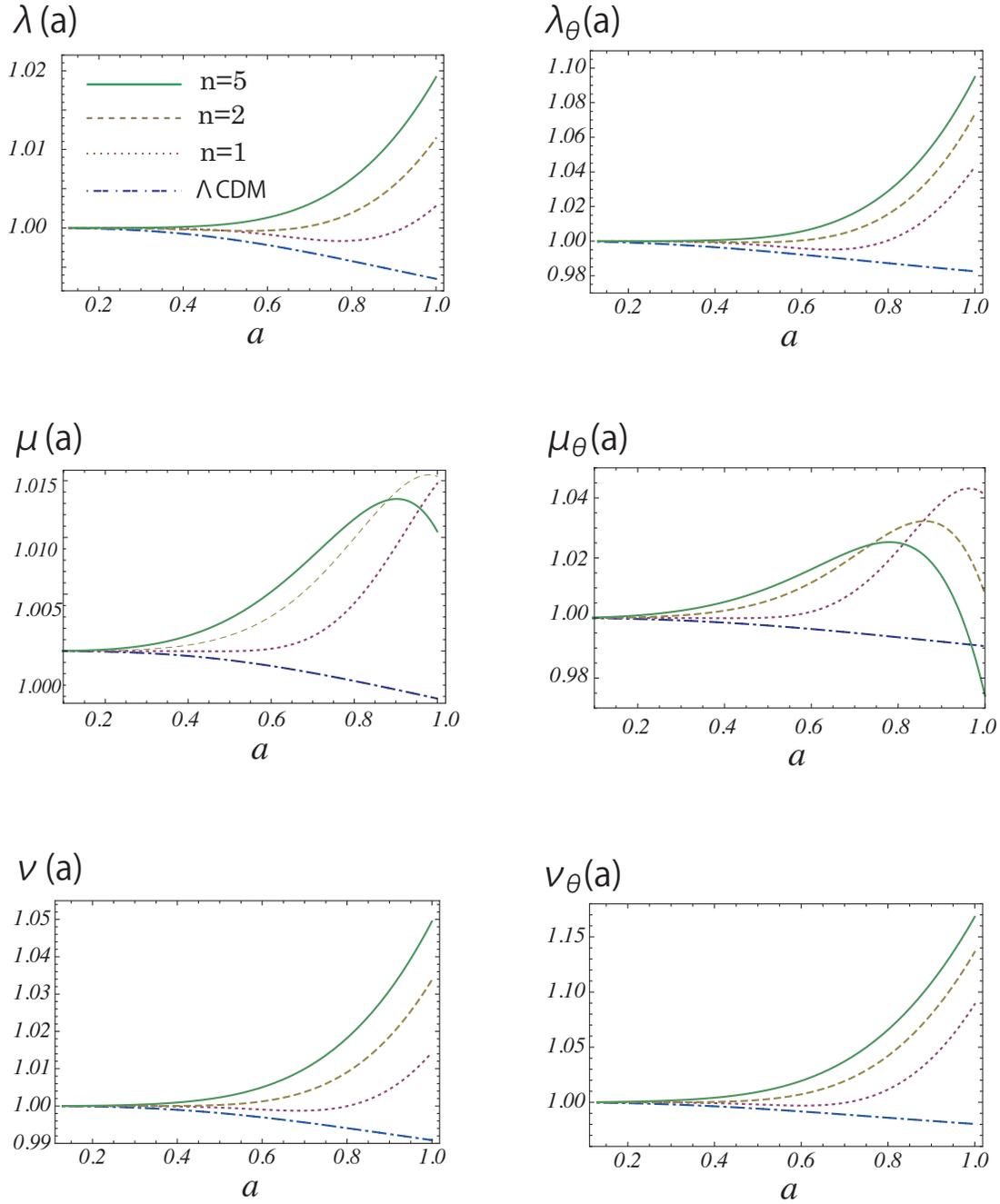}
  \end{center}
 \caption{$\lambda, ~\mu, ~\nu,  ~\lambda_\theta, ~ \mu_\theta, ~ \nu_\theta$ as function of the scale factor $a$.
In each panel, the blue dash-dotted curve is the $\Lambda {\rm CDM}$ model, and the red dotted curve, the yellow dashed curve, 
and the green thick solid curve are the KGB model with $n=1$, ~$2$, and $5$, respectively. 
\label{fig:six}
}
\end{figure}


\begin{figure}[]
  \begin{center}
   \hspace{0mm}\scalebox{.7}{\includegraphics{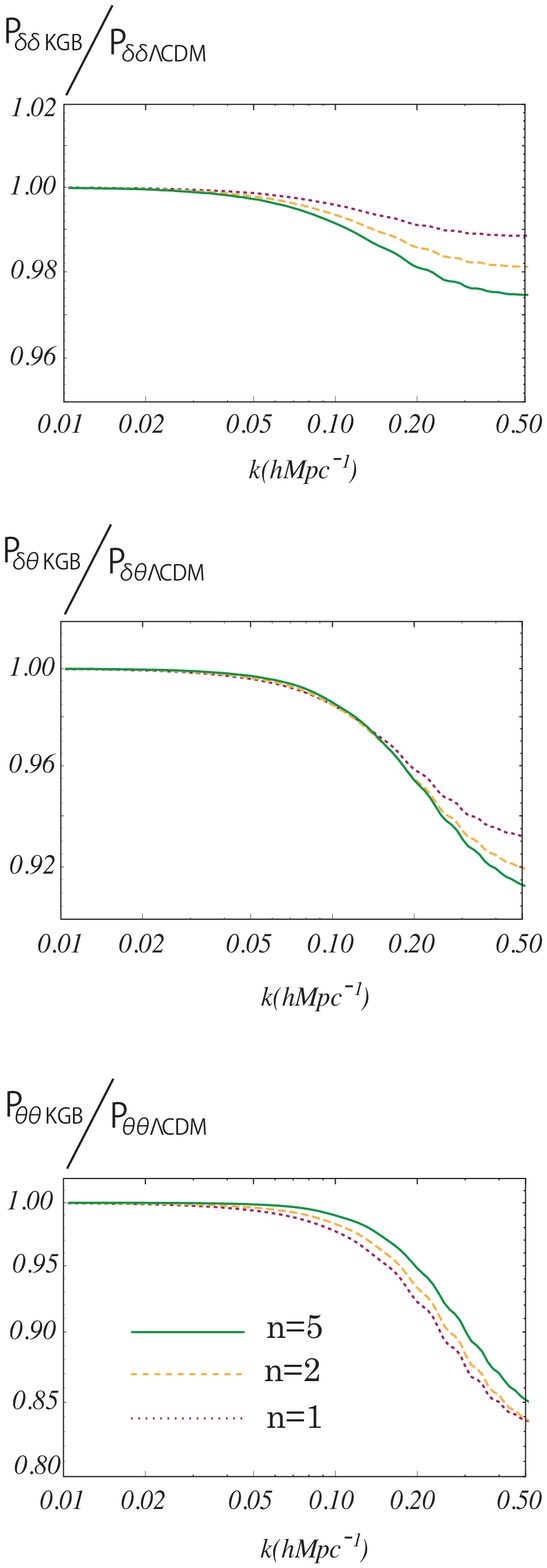}}
  \end{center}
 \caption{
Relative deviation of the power spectra $P_{\delta\delta}(k)$ (top panel), ~$P_{\delta\theta}(k)$ (middle pane.), ~$P_{\theta\theta}(k)$ (bottom panel), under the kinetic gravity braiding model with $n = 1$ (red dotted curve), $n = 2$ (yellow dashed curve), $n = 5$ (green thick curve), which are divided by those under the $\Lambda {\rm CDM}$ model.
The panels show the snapshot at the redshift $z=0$. 
\label{fig:sixx}
}
\end{figure}

Figure \ref{fig:six} shows $\lambda, ~\mu, ~\nu,  ~\lambda_\theta, ~ \mu_\theta, ~ \nu_\theta$ as function of the scale factor $a$.
In each panel, the blue dash-dotted curve is the $\Lambda {\rm CDM}$ model, and the red dotted curve, the yellow dashed curve, 
and the green thick solid curve are the KGB model with $n=1$, ~$2$, and $5$, respectively. 
All the curves take the limiting value unity at $a=0$, 
but deviate from the unity as $a$ evolves.  
Note that the deviation of $\lambda,  ~\mu,  ~\nu$ from unity is small, of the order of a few percent,
but the deviation of $\lambda_\theta, ~\nu_\theta$ is rather large, which could be $10$ percent.
This is because the parameters associated with the velocity, $\lambda_\theta$ and $\nu_\theta$ 
defined by Eqs.~(\ref{lambdatheta}) and (\ref{nutheta}), respectively, contain the time 
derivative term, which makes a large contribution. 
Plus, some part of the difference between the $\Lambda {\rm CDM}$ and the KGB model come from the 
difference of the growth rate $f$. 
Deviation of $\mu_\theta$ in the KGB model from that in the $\Lambda$CDM model 
is rather small compared with the deviations of $\lambda_\theta$ and $\nu_\theta$, 
which comes from the fact that $\mu$ is not a monotonic increasing function 
but there exists a maximum value at $a\simlt1$.

Figure \ref{fig:sixx} shows the 1-loop power spectra $P_{\delta\delta}$, ~$P_{\delta\theta}$, ~$P_{\theta\theta}$, from the top to the bottom, respectively, which are 
normalized by those of the $\Lambda {\rm CDM}$ model. 
These are the snapshots at $z=0$, and we adopted the same normalization begin $\sigma_8=0.8$ 
for each model, which means that all the models have the same linear matter power spectrum. 
In each panel, the red dotted curve, the yellow dashed curve, and the green thick curve 
show the KGB model with $n=1$, $2$ and $5$, respectively. 
In the linear regime $k\simlt 0.1[h{\rm Mpc}^{-1}]$,  
all the models converge because they have the same linear matter power spectrum
due to the same normalization  $\sigma_8=0.8$. 
The differences between the KGB model and the $\Lambda$CDM model 
appear for the quasi-nonlinear regime 
$k\simgt 0.1[h{\rm Mpc}^{-1}]$ due to the nonlinear effect. 
Because all the model have the same linear matter power spectrum, this figure
shows that the enhancement of the power spectrum due to the nonlinear effect 
is small in the KGB model compared with that in the $\Lambda {\rm CDM}$ model. 
This is understood as the results of the Vainshtein effect. 
Furthermore, the deviation from the $\Lambda {\rm CDM}$ model is more significant in 
the velocity power spectrum than that in the density power spectrum. 
In general, the amplitude of the 1-loop power spectra $P_{\delta\delta}$, ~$P_{\delta\theta}$, 
and $P_{\theta\theta}$ are decreased when any of $\lambda(t)$, $\mu(t)$, 
$\lambda_{\theta}(t)$, and $\mu_\theta(t)$ is increased. 
The behavior of $P_{\delta\theta}$ and $P_{\theta\theta}$ in the quasi-nonlinear regime 
is dominantly influenced by $\lambda_\theta(t)$ and $\mu_\theta(t)$.

\section{Summary and conclusions}
We found the third order solutions of the cosmological density perturbations
in the Horndeski's most general scalar-tensor theory assuming that the 
Vainshtein mechanism is at work. 
We solved the equations under the quasi-static approximation, and
the solutions describe the quasi-nonlinear aspects of the cosmological 
density contrast and the velocity divergence under the Vainshtein mechanism. 
In this work, we thoroughly investigate the independence property of 
the mode-couplings functions describing the non-linear interactions. 
We found that the third order solution of the density contrast is 
characterized by $3$ parameters for the nonlinear interactions, one 
of which is the same as that for the second order solutions. 
The third order solution of the velocity divergence is characterized by 
$4$ parameters for the nonlinear interactions, two of which 
are the same parameters as those of the second order solutions. 
The nonlinear features of the perturbative solutions up to the third order 
are characterized by $6$ parameters. Furthermore, the 1-loop order power spectra 
obtained with the third order solutions are described by 4 parameters. 
Assuming the KGB model, we demonstrated the effect of the modified gravity in the 
1-loop order power spectra at the quantitative level.  
We found that the deviation from the $\Lambda {\rm CDM}$ model appears in the
power spectra of the density contrast and the velocity divergence, which can
be understood as the results of the Vainshtein mechanism. 
The deviation from the $\Lambda {\rm CDM}$ model is more significant
in the velocity divergence than the density contrast, which 
is explained by a dominant contribution of the parameters $\lambda_{\theta}(t)$ and $\mu_\theta(t)$. 
It is interesting to investigate whether this is a general 
feature of the modified gravity with the Vainshtein mechanism or not.

\section*{Acknowledgment}
This work is supported by a research support program of Hiroshima University.

\appendix
\section{Coefficients in equations and solutions}
We summarize the definitions of the coefficients in the gravitational and scalar 
field equations (\ref{trlseq}), (\ref{00eq}), (\ref{seom}).
\begin{eqnarray}
A_0&=&\frac{\dot\Theta}{H^2}+\frac{\Theta}{H}
+{\cal F}_T-2{\cal G}_T-2\frac{\dot{\cal G}_T}{H}-\frac{{\cal E}+{\cal P}}{2H^2},
\\
A_1&=&\frac{\dot{\cal G}_T}{H}+ {\cal G}_T-{\cal F}_T,
\\
A_2&=& {\cal G}_T-\frac{\Theta}{H},
\\
B_0&=&\frac{X}{H}\biggl\{\dot\phi G_{3X}+3\left(\dot X+2HX\right)G_{4XX}
+2X\dot XG_{4XXX}-3\dot\phi G_{4\phi X}
+2\dot\phi XG_{4\phi XX}
\nonumber\\&&
+\left(\dot H+H^2\right)\dot\phi G_{5X}
+\dot\phi
\left[2H\dot X+\left(\dot H+H^2\right) X\right]G_{5XX}
+H\dot\phi X\dot XG_{5XXX}
\nonumber\\&&
-2\left(\dot X+2HX\right)G_{5\phi X}
-\dot\phi XG_{5\phi\phi X}-X\left(\dot X-2HX\right)G_{5\phi XX}\biggr\},
\\
B_1&=&2X\left[G_{4X}+\ddot\phi\left(G_{5X}+XG_{5XX}\right)
-G_{5\phi}+XG_{5\phi X}\right],
\\
B_2&=&
-2X\left(G_{4X}+2XG_{4XX}+H\dot\phi G_{5X}
+H\dot\phi XG_{5XX}-G_{5\phi}-XG_{5\phi X}\right),
\\
B_3&=&H\dot\phi XG_{5X},
\\
C_0&=&2X^2G_{4XX}+\frac{2X^2}{3}\left(2\ddot\phi G_{5XX}
+\ddot\phi XG_{5XXX}-2G_{5\phi X}+XG_{5\phi XX}\right),
\\
C_1&=&H\dot\phi X\left(G_{5X}+XG_{5XX}\right),
\end{eqnarray}
where we define 
\begin{eqnarray}
{\cal F}_T&=&2\left[G_4
-X\left( \ddot\phi G_{5X}+G_{5\phi}\right)\right],
\\
{\cal G}_T&=&2\left[G_4-2 XG_{4X}
-X\left(H\dot\phi G_{5X} -G_{5\phi}\right)\right],
\\
\Theta&=&-\dot\phi XG_{3X}+
2HG_4-8HXG_{4X}
-8HX^2G_{4XX}+\dot\phi G_{4\phi}+2X\dot\phi G_{4\phi X}
\nonumber\\&&
-H^2\dot\phi\left(5XG_{5X}+2X^2G_{5XX}\right)
+2HX\left(3G_{5\phi}+2XG_{5\phi X}\right),
\\
{\cal E} &=& 2 X K_X - K + 6 X \dot \phi  H G_{3X} - 2 X G_{3 \phi} - 6 H^2 G_4
 + 24 H^2 X (G_{4X} + X G_{4XX})
\nonumber\\&&
 - 12 H X \dot \phi G_{4 \phi X}- 6 H \dot \phi  G_{4\phi} + 2 H^3 X \dot \phi (5 G_{5 X} 
 + 2 X G_{5XX}) 
\nonumber\\&&
- 6 H^2 X (3 G_{5 \phi} + 2 X G_{5\phi X}),
\\
{\cal P}&=& K - 2X(G_{3 \phi} + \ddot \phi G_{3 X}) + 2(3 H^2 + 2 \dot H)G_4 
- 12 H^2 X G_{4X} - 4 H \dot X G_{4 X} 
\nonumber\\&&
- 8 \dot H X G_{4X} - 8 H X \dot X G_{4 X X} 
+ 2 (\ddot \phi + 2 H \dot \phi)G_{4 \phi} + 4 X G_{4 \phi \phi} + 4 X (\ddot \phi
 - 2 H \dot \phi)G_{4\phi X} 
\nonumber\\&&
- 2 X (2 H^3 \dot \phi + 2 H \dot H \dot \phi + 3 H^2 \ddot \phi)G_{5 X}
- 4 H^2 X^2 \ddot \phi G_{5XX} + 4 H X (\dot X - H X)G_{5 \phi X} 
\nonumber\\&&+ 2\left[ 2 (H X) \dot{}
 + 3 H^2 X \right]G_{5\phi} + 4 H X \dot \phi G_{5 \phi \phi}.
\end{eqnarray}
The coefficients in the first and the second order solutions are defined as follows,
\begin{eqnarray}
{\cal R}(t) &=& A_0{\cal F}_T - A_1^2,\\
{\cal S}(t) &=& A_0{\cal G}_T + A_1 A_2,\\
{\cal T}(t) &=& A_1 {\cal G}_T + A_2 {\cal F}_T,\\
{\cal Z}(t) &=& 2\left(A_0 {\cal G}_T^2 + 2 A_1 A_2 {\cal G}_T + A_2^2 {\cal F}_T\right),\\
\kappa_\Phi(t) &=& {\rho_{\rm m} {\cal R}\over H^2 {\cal Z}},\\
\kappa_\Psi(t) &=& {\rho_{\rm m} {\cal S}\over H^2 {\cal Z}},\\
\kappa_Q(t) &=& {\rho_{\rm m} {\cal T}\over H^2 {\cal Z}},\\
N_\gamma (t) &=& {H^4 \over \rho_{\rm m}} \left(
2 B_0 \kappa_Q^3 - 3 B_1\kappa_\Psi \kappa_Q^2 -3 B_2 \kappa_\Phi \kappa_Q^2 -6 B_3 \kappa_\Phi \kappa_\Psi \kappa_Q \right),
\label{NGAMMA}
\\
\lambda(t) &=& {7\over 2D_+^2(t)}\int_0^t{D_-(t)D_+(t')-D_+(t)D_-(t')\over W(t')}D_+^2(t')\left(f^2H^2-N_\gamma(t')\right)dt', \label{2ndlambda}\\
\lambda_{\theta}(t) &=& \lambda(t) + {\dot{\lambda}(t) \over 2fH}, \label{lambdatheta}\\
\lambda_{\Phi}(t) &=& - {2\over 7}\kappa_\Phi\lambda(t)
+  {1 \over {\cal Z}} \left(2 B_0 {\cal T} \kappa_Q^2
- 3 B_1{\cal S} \kappa_Q^2 -3 B_2 {\cal R} \kappa_Q^2 - 6 B_3 {\cal R} \kappa_\Psi \kappa_Q\right),\\
\lambda_{\Psi}(t) &=& - {2\over 7}\kappa_\Psi\lambda(t)
+ {1\over {\cal Z} } \left(2 B_0 A_2{\cal G}_T \kappa_Q^2 
+  B_1(A_2^2 \kappa_Q^2 - 2 A_2 {\cal G}_T \kappa_\Psi \kappa_Q) -B_2({\cal S} \kappa_Q^2 + 2 A_2 {\cal G}_T \kappa_\Phi \kappa_Q)\right. 
\nonumber\\
&&\hspace{2.3cm}\left.
- 2 B_3({\cal S} \kappa_\Psi \kappa_Q - A_2^2 \kappa_\Phi \kappa_Q + A_2 {\cal G}_T \kappa_\Phi \kappa_\Psi)
\right),
\\
\lambda_{Q}(t) &=& - {2\over 7}\kappa_Q\lambda(t)
- {1\over {\cal Z}} \left(2 B_0 {\cal G}_T^2 \kappa_Q^2
+ B_1(A_2 {\cal G}_T \kappa_Q^2 - 2 {\cal G}_T^2 \kappa_\Psi \kappa_Q)+ B_2({\cal T}\kappa_Q^2 - 2 {\cal G}_T^2 \kappa_\Phi \kappa_Q ) \right.
\nonumber\\
&&\hspace{2.3cm}\left.
+ 2 B_3({\cal T} \kappa_\Psi \kappa_Q + A_2 {\cal G}_T \kappa_\Phi \kappa_Q  - {\cal G}_T^2 \kappa_\Phi \kappa_\Psi )
\right).
\end{eqnarray}
Some details are described in the previous paper \cite{TTY}, but one can show that
$\lambda(t)$ obeys the differential equation,
\begin{eqnarray}
\ddot{\lambda}(t) + (4 f + 2)H \dot{\lambda}(t) + (2 f^2 H^2 - L) \lambda(t) = {7\over 2}(f^2 H^2 - N_\gamma). 
\label{2ndlambdade}
\end{eqnarray}

The coefficients for the third order solutions are defined as
\begin{eqnarray}
\sigma_\Phi(t) &=&{2 \over {\cal Z}} \left(2B_0 {\cal T}\kappa_Q^2 - 3 B_1{\cal S} \kappa_Q^2 
- 3 B_2{\cal R} \kappa_Q^2 - 6 B_3 {\cal R} \kappa_\Psi \kappa_Q\right),
\label{SIGMAPHI}\\
\mu_\Phi(t) &=& {2 \over {\cal Z}} \left(2 B_0 {\cal T}\kappa_Q \lambda_{Q} - B_1\left(2 {\cal S} \kappa_Q \lambda_{Q} +  {\cal T} \kappa_Q \lambda_{\Psi} \right) - B_2 \left(2 {\cal R} \kappa_Q \lambda_Q +  {\cal T}\kappa_Q \lambda_\Phi\right) \right.\nonumber\\
&&\hspace{2.49cm} \left.- 2 B_3\left({\cal R} \kappa_\Psi \lambda_Q + {\cal R} \kappa_Q \lambda_\Psi + {\cal S} \kappa_Q \lambda_\Phi \right)\right),\\
\nu_\Phi(t) &=& {2 \over 3 {\cal Z}} \left(- 3C_0 {\cal T} \kappa_Q^3 - 4C_1{\cal R} \kappa_Q^3 \right),\\
\sigma_\Psi(t) &=& {2 \over {\cal Z}} \left( 2 B_0 A_2{\cal G}_T \kappa_Q^2 - B_1\left( 2A_2 {\cal G}_T \kappa_\Psi \kappa_Q - A_2^2 \kappa_Q^2 \right) - B_2 \left({\cal S} \kappa_Q^2 + 2 A_2 {\cal G}_T \kappa_\Phi \kappa_Q \right) \right.\nonumber\\
&&\hspace{2.49cm} 
\left.- 2 B_3\left({\cal S} \kappa_\Psi \kappa_Q + A_2 {\cal G}_T \kappa_\Phi \kappa_\Psi - A_2^2 \kappa_\Phi \kappa_Q \right)\right),\\
\mu_\Psi(t) &=& {2\over {\cal Z}} \left(2 B_0 A_2{\cal G}_T \kappa_Q \lambda_Q
- B_1\left(A_2 {\cal G}_T \left(\kappa_\Psi \lambda_Q + \kappa_Q \lambda_\Psi\right) -  A_2^2 \kappa_Q \lambda_Q \right) \right. \nonumber\\
&&\hspace{1.0cm} 
- B_2 \left({\cal S} \kappa_Q \lambda_Q + A_2 {\cal G}_T \left(\kappa_\Phi \lambda_Q + \kappa_Q \lambda_\Phi\right) \right) \nonumber\\
&&\hspace{1.0cm} 
\left.- B_3\left({\cal S}\left( \kappa_\Psi \lambda_Q + \kappa_Q \lambda_{\Psi}\right) + A_2 {\cal G}_T \left(\kappa_\Phi \lambda_\Psi + \kappa_\Psi \lambda_\Phi\right) - A_2^2\left(\kappa_\Phi \lambda_Q + \kappa_Q \lambda_\Phi\right) \right)\right),\\
\nu_\Psi(t) &=& {2\over 3 {\cal Z}} \left(- 3 C_0 A_2 {\cal G}_T \kappa_Q^3  - C_1 \left({\cal S} \kappa_Q^3 + 3 A_2 {\cal G}_T \kappa_\Phi \kappa_Q^2 \right)\right),\\
\sigma_Q(t) &=& {2\over {\cal Z}} \left(- 2 B_0 {\cal G}_T^2 \kappa_Q^2 
+ B_1\left(2 {\cal G}_T^2 \kappa_\Psi \kappa_Q - A_2 {\cal G}_T \kappa_Q^2 \right)
- B_2 \left({\cal T}\kappa_Q^2 - 2 {\cal G}_T^2\kappa_\Phi \kappa_Q \right)\right. \nonumber\\
&&\hspace{1.0cm} 
\left.- 2 B_3\left({\cal S} \kappa_Q^2 - {\cal G}_T^2 \kappa_\Phi \kappa_\Psi + A_2 {\cal G}_T \kappa_\Phi \kappa_Q \right)\right),\\
\mu_Q(t) &=& {2\over {\cal Z}} \left(- 2 B_0 {\cal G}_T^2 \kappa_Q \lambda_Q
+ B_1\left({\cal G}_T^2 \left(\kappa_\Psi \lambda_Q + \kappa_Q \lambda_\Psi\right) - A_2 {\cal G}_T \kappa_Q\lambda_Q\right) \right.\nonumber\\
&&\hspace{0.9cm} 
- B_2 \left({\cal T} \kappa_Q \lambda_Q - {\cal G}_T^2\left(\kappa_\Phi \lambda_Q + \kappa_Q \lambda_\Phi\right) \right) \nonumber\\
&&\hspace{0.9cm} 
\left.- B_3\left(\left({\cal S} \kappa_Q \lambda_Q + {\cal T} \kappa_Q \lambda_\Psi\right) - {\cal G}_T^2 \left(\kappa_\Phi \lambda_\Psi + \kappa_\Psi \lambda_{\Phi}\right) + A_2 {\cal G}_T \left(\kappa_\Phi \lambda_Q + \kappa_Q \lambda_\Phi\right) \right)\right),\\
\nu_Q(t) &=& {2\over 3 {\cal Z}} \left(3 C_0 {\cal G}_T^2 \kappa_Q^3 +  C_1 \left(- {\cal T} \kappa_Q^3 + 3 {\cal G}_T^2 \kappa_\Phi \kappa_Q^2 \right)\right).
\end{eqnarray}

\section{The third order mode-coupling functions}
In this appendix, we summarize the functions that describe the 
nonlinear mode-couplings of the third order solutions. In order to derive
Eqs.~(\ref{GRAA}), (\ref{GRAB}) and (\ref{GRAC}), we define
\begin{eqnarray}
{\cal W}_{\gamma \alpha}({\bf p})&\equiv&{1\over (2 \pi)^6}\int d{\bf k}_1 d{\bf k}_2 d{\bf k}_3\delta^{(3)}({\bf k}_1+{\bf k}_2+{\bf k}_3-{\bf p})\gamma\alpha({\bf k}_1,{\bf k}_2,{\bf k}_3) \delta_L({\bf k}_1)\delta_L({\bf k}_2)\delta_L({\bf k}_3),
\label{WGA}
\\
{\cal W}_{\gamma\gamma}({\bf p})&\equiv&{1\over (2 \pi)^6}\int d{\bf k}_1 d{\bf k}_2 d{\bf k}_3\delta^{(3)}({\bf k}_1+{\bf k}_2+{\bf k}_3-{\bf p})\gamma\gamma({\bf k}_1,{\bf k}_2,{\bf k}_3)\delta_L({\bf k}_1)\delta_L({\bf k}_2)\delta_L({\bf k}_3),
\label{WGG}
\\
{\cal W}_\xi({\bf p})&\equiv&{1\over (2 \pi)^6}\int d{\bf k}_1d{\bf k}_2 d{\bf k}_{3}  \delta^{(3)}({\bf k}_1 + {\bf k}_2 + {\bf k}_3 - {\bf p})\xi({\bf k}_1,{\bf k}_2,{\bf k}_3) \delta_L({\bf k}_1)\delta_L({\bf k}_2 )\delta_L({\bf k}_3),
\label{WXI}
\end{eqnarray}
with
\begin{eqnarray}
&&\gamma\alpha({\bf k}_1,{\bf k}_2,{\bf k}_3)
={1\over3}\left( \gamma({\bf k}_1,{\bf k}_2 + {\bf k}_3)\alpha^{(s)}({\bf k}_2,{\bf k}_3)+{\rm 2~cyclic~terms}\right), 
\\
&&\gamma\gamma({\bf k}_1,{\bf k}_2,{\bf k}_3)
={1\over3}\left( \gamma({\bf k}_1,{\bf k}_2 + {\bf k}_3)\gamma({\bf k}_2,{\bf k}_3)+{\rm 2~cyclic~terms}\right), 
\\
&&\xi({\bf k}_1,{\bf k}_2,{\bf k}_3)=1 - {k_1^2({\bf k}_2\cdot{\bf k}_3)^2 + k_2^2({\bf k}_3\cdot{\bf k}_1)^2 + k_3^2({\bf k}_1\cdot{\bf k}_2)^2 - 2({\bf k}_1\cdot{\bf k}_2)({\bf k}_2\cdot{\bf k}_3)({\bf k}_3\cdot{\bf k}_1)\over k_1^2 k_2^2 k_3^2}.
\end{eqnarray}
In deriving Eqs.~(\ref{eoc33}) and (\ref{eoe33}), we define 
\begin{eqnarray}
{\cal W}_{\alpha \alpha R}({\bf p})&=&{1\over (2 \pi)^6}\int d{\bf k}_1 d{\bf k}_2 d{\bf k}_3\delta^{(3)}({\bf k}_1+{\bf k}_2+{\bf k}_3-{\bf p})\alpha\alpha_R({\bf k}_1,{\bf k}_2,{\bf k}_3) \delta_L({\bf k}_1)\delta_L({\bf k}_2)\delta_L({\bf k}_3),
\label{WAAR}
\\
{\cal W}_{\alpha\gamma R}({\bf p})&=&{1\over (2 \pi)^6}\int d{\bf k}_1 d{\bf k}_2 d{\bf k}_3\delta^{(3)}({\bf k}_1+{\bf k}_2+{\bf k}_3-{\bf p})\alpha\gamma_R({\bf k}_1,{\bf k}_2,{\bf k}_3) \delta_L({\bf k}_1)\delta_L({\bf k}_2)\delta_L({\bf k}_3),\\
{\cal W}_{\alpha \alpha L}({\bf p})&=&{1\over (2 \pi)^6}\int d{\bf k}_1 d{\bf k}_2 d{\bf k}_3\delta^{(3)}({\bf k}_1+{\bf k}_2+{\bf k}_3-{\bf p})\alpha\alpha_L({\bf k}_1,{\bf k}_2,{\bf k}_3) \delta_L({\bf k}_1)\delta_L({\bf k}_2)\delta_L({\bf k}_3),\\
{\cal W}_{\alpha\gamma L}({\bf p})&=&{1\over (2 \pi)^6}\int d{\bf k}_1 d{\bf k}_2 d{\bf k}_3\delta^{(3)}({\bf k}_1+{\bf k}_2+{\bf k}_3-{\bf p})\alpha\gamma_L({\bf k}_1,{\bf k}_2,{\bf k}_3) \delta_L({\bf k}_1)\delta_L({\bf k}_2)\delta_L({\bf k}_3),\\
{\cal W}_{\alpha \alpha}({\bf p})&=&{1\over (2 \pi)^6}\int d{\bf k}_1 d{\bf k}_2 d{\bf k}_3\delta^{(3)}({\bf k}_1+{\bf k}_2+{\bf k}_3-{\bf p})\alpha\alpha({\bf k}_1,{\bf k}_2,{\bf k}_3) \delta_L({\bf k}_1)\delta_L({\bf k}_2)\delta_L({\bf k}_3),\\
{\cal W}_{\alpha\gamma}({\bf p})&=&{1\over (2 \pi)^6}\int d{\bf k}_1 d{\bf k}_2 d{\bf k}_3\delta^{(3)}({\bf k}_1+{\bf k}_2+{\bf k}_3-{\bf p})\alpha\gamma({\bf k}_1,{\bf k}_2,{\bf k}_3) \delta_L({\bf k}_1)\delta_L({\bf k}_2)\delta_L({\bf k}_3)
\label{WAR}
\end{eqnarray}
with
\begin{eqnarray}
\alpha\alpha_R({\bf k}_1,{\bf k}_2,{\bf k}_3)&=&
{1\over3}\left(\alpha({\bf k}_1,{\bf k}_2 + {\bf k}_3)\alpha^{(s)}({\bf k}_2,{\bf k}_3) +{\rm 2~cyclic~terms}\right), 
\\
\alpha\gamma_R({\bf k}_1,{\bf k}_2,{\bf k}_3)&=&
{1\over3}\left(\alpha({\bf k}_1,{\bf k}_2 + {\bf k}_3)\gamma({\bf k}_2,{\bf k}_3) +{\rm 2~cyclic~terms}\right), 
\\
\alpha\alpha_L({\bf k}_1,{\bf k}_2,{\bf k}_3)&=&
{1\over3}\left(\alpha({\bf k}_1 + {\bf k}_2, {\bf k}_3)\alpha^{(s)}({\bf k}_2,{\bf k}_3) +{\rm 2~cyclic~terms}\right), 
\\
\alpha\gamma_L({\bf k}_1,{\bf k}_2,{\bf k}_3)&=&
{1\over3}\left(\alpha({\bf k}_1 + {\bf k}_2, {\bf k}_3)\gamma({\bf k}_2,{\bf k}_3) +{\rm 2~cyclic~terms}\right), 
\\
\alpha\alpha({\bf k}_1,{\bf k}_2,{\bf k}_3)&=&
{1\over3}\left(\alpha^{(s)}({\bf k}_1,{\bf k}_2 + {\bf k}_3)\alpha^{(s)}({\bf k}_2,{\bf k}_3) +{\rm 2~cyclic~terms}\right),
\\
\alpha\gamma({\bf k}_1,{\bf k}_2,{\bf k}_3)&=&
{1\over3}\left(\alpha^{(s)}({\bf k}_1,{\bf k}_2 + {\bf k}_3)\gamma({\bf k}_2,{\bf k}_3) +{\rm 2~cyclic~terms}\right).
\end{eqnarray}

\section{Derivation of the 1-loop power spectra} 
The cosmological density contrast $\delta(t, {\bf k})$ and the velocity divergence 
$\theta(t, {\bf k})$ 
up to the third order of the perturbative expansion are expressed as
\begin{eqnarray}
\delta(t, {\bf k}) &=& D_+(t) \delta_{\rm L}({\bf k}) + D_+^2(t) \delta_{2K}(t, {\bf k}) + D_+^3(t) \delta_{3K}(t, {\bf k}),\label{delta}\\
\theta(t, {\bf k}) &=& - f \left(D_+(t) \delta_{\rm L}({\bf k}) + D_+^2(t) \theta_{2K}(t, {\bf k}) + D_+^3(t) \theta_{3K}(t, {\bf k}) \right),\label{theta}
\end{eqnarray}
were we define 
\begin{eqnarray}
\delta_{2K}(t, {\bf k}) &=& {\cal W}_\alpha({\bf k}) - {2\over 7}\lambda(t){\cal W}_\gamma({\bf k}),\label{d2k}\\
\delta_{3K}(t, {\bf k}) &=& {\cal W}_{\alpha\alpha}({\bf k}) - {2\over 7}\lambda(t){\cal W}_{\alpha\gamma R}({\bf k}) - {2\over 7}\lambda(t){\cal W}_{\alpha\gamma L}({\bf k}) - {2\over 21}\mu(t){\cal W}_{\gamma\gamma}({\bf k}) + {1\over 9}\nu(t){\cal W}_\xi({\bf k}),\label{d3k}\\
\theta_{2K}(t, {\bf k}) &=& {\cal W}_\alpha({\bf k}) - {4\over 7}\lambda_\theta(t){\cal W}_\gamma({\bf k}),\label{t2k}\\
\theta_{3K}(t, {\bf k}) &=& {\cal W}_{\alpha\alpha}({\bf k}) - {4\over 7}\lambda_\theta(t){\cal W}_{\alpha\gamma R}({\bf k}) - {2\over 7}\lambda(t){\cal W}_{\alpha\gamma L}({\bf k}) - {2\over 7}\mu_\theta(t){\cal W}_{\gamma\gamma}({\bf k}) + {1\over 3}\nu_\theta(t){\cal W}_\xi({\bf k}),\label{t3k}
\end{eqnarray}
and the kernels for the density contrast $F_2$, and $F_3$, and those for the velocity divergence 
$G_2$, and $G_3$, as follows, 
\begin{eqnarray}
&&F_2(t,{\bf k}_1, {\bf k}_2) = \alpha({\bf k}_1, {\bf k}_2) - {2\over 7}\lambda(t)\gamma({\bf k}_1, {\bf k}_2),
\label{efu2}
\\
&&G_2(t,{\bf k}_1, {\bf k}_2) = \alpha({\bf k}_1, {\bf k}_2) - {4\over 7}\lambda_\theta(t)\gamma({\bf k}_1, {\bf k}_2),\\
&&F_3(t,{\bf k}_1, {\bf k}_2, {\bf k}_3) = \alpha\alpha({\bf k}_1, {\bf k}_2, {\bf k}_3) - {2\over 7}\lambda(t)\alpha\gamma_R({\bf k}_1, {\bf k}_2, {\bf k}_3) - {2\over 7}\lambda(t)\alpha\gamma_L({\bf k}_1, {\bf k}_2, {\bf k}_3)
\nonumber\\
&&\hspace{2.2cm}- {2\over 21}\mu(t)\gamma\gamma({\bf k}_1, {\bf k}_2, {\bf k}_3)+ {1\over 9}\nu(t)\xi({\bf k}_1, {\bf k}_2, {\bf k}_3),
\label{efu3}\\
&&G_3(t,{\bf k}_1, {\bf k}_2, {\bf k}_3) = \alpha\alpha({\bf k}_1, {\bf k}_2, {\bf k}_3) - {4\over 7}\lambda_\theta(t)\alpha\gamma_R({\bf k}_1, {\bf k}_2, {\bf k}_3) - {2\over 7}\lambda(t)\alpha\gamma_L({\bf k}_1, {\bf k}_2, {\bf k}_3) \nonumber\\
&&\hspace{2.2cm}- {2\over 7}\mu_\theta(t)\gamma\gamma({\bf k}_1, {\bf k}_2, {\bf k}_3)+ {1\over 3}\nu_\theta(t)\xi({\bf k}_1, {\bf k}_2, {\bf k}_3).
\end{eqnarray}
These kernels have the two types of symmetries. One is the symmetries
in replacement of the wave numbers,  
\begin{eqnarray}
&&F_2(t, {\bf k}_1, {\bf k}_2) = F_2(t, {\bf k}_2, {\bf k}_1),\\
\label{efu2sym}
&&F_3(t, {\bf k}_1, {\bf k}_2, {\bf k}_3) = F_3(t, {\bf k}_2, {\bf k}_3, {\bf k}_1) = F_3(t, {\bf k}_3, {\bf k}_1, {\bf k}_2)\nonumber\\
&&\hspace{2.48cm} = F_3(t, {\bf k}_1, {\bf k}_3, {\bf k}_2) = F_3(t, {\bf k}_2, {\bf k}_1, {\bf k}_3) = F_3(t, {\bf k}_3, {\bf k}_2, {\bf k}_1). 
\label{efu3sym}
\end{eqnarray}  
The second is the symmetries in the conversion of the sign of the wavenumbers,
\begin{eqnarray}
F_2(t, {\bf k}_1, {\bf k}_2) &=& F_2(t, - {\bf k}_1, - {\bf k}_2),\label{f2ss}\\
F_3(t, {\bf k}_1, {\bf k}_2, {\bf k}_3) &=& F_3(t, - {\bf k}_1, - {\bf k}_2, - {\bf k}_3).\label{f3ss}
\end{eqnarray}
The same relations hold for $G_2(t, {\bf k}_1, {\bf k}_2)$ and $G_3(t, {\bf k}_1, {\bf k}_2, {\bf k}_3)$. 

The above properties are useful in deriving the expressions of the power spectra, 
$P_{\delta\delta}(t, k)$, ~ $P_{\delta\theta}(t, k)$,~$P_{\theta\theta}(t, k)$, 
defined by Eqs.~(\ref{definedd}), (\ref{definedt}) and (\ref{definett}).
Using the expressions (\ref{delta}) and (\ref{theta}), 
we find 
\begin{eqnarray}
P_{\delta\delta}(t, k) &=& D_+^2(t) P_{\rm L}(k) + D_+^4(t)\left(P_{\delta\delta}^{(22)}(t, k) + 2 P_{\delta\delta}^{(13)}(t, k)\right),\\
P_{\delta\theta}(t, k) &=& D_+^2(t) P_{\rm L}(k) + D_+^4(t)\left(P_{\delta\theta}^{(22)}(t, k) + 2 P_{\delta\theta}^{(13)}(t, k)\right),\\
P_{\theta\theta}(t, k) &=& D_+^2(t) P_{\rm L}(k) + D_+^4(t)\left(P_{\theta\theta}^{(22)}(t, k) + 2 P_{\theta\theta}^{(13)}(t, k)\right),
\end{eqnarray}
where $D_+^2(t)P_{\rm L}(k)$ is the linear matter power spectrum, 
\begin{eqnarray}
\left<\delta_{\rm L}({\bf k}_1) \delta_{\rm L}({\bf k}_2)\right> &=& (2\pi)^3 \delta^{(3)}({\bf k}_1+{\bf k}_2)P_{\rm L}(k), \label{lps}
\end{eqnarray}
and we define 
\begin{eqnarray}
\left<\delta_{2K}(t, {\bf k}_1) \delta_{2K}(t, {\bf k}_2)\right> &=& (2\pi)^3 \delta^{(3)}({\bf k}_1+{\bf k}_2)P_{\delta\delta}^{(22)}(t, k), \label{ps22dd}\\
\left<\delta_{\rm L}({\bf k}_1) \delta_{3K}(t, {\bf k}_2)\right> &=& (2\pi)^3 \delta^{(3)}({\bf k}_1+{\bf k}_2)P_{\delta\delta}^{(13)}(t, k), \label{ps13dd}\\
\left<\delta_{2K}(t, {\bf k}_1) \theta_{2K}(t, {\bf k}_2)\right> &=& (2\pi)^3 \delta^{(3)}({\bf k}_1+{\bf k}_2)P_{\delta\theta}^{(22)}(t, k), \label{ps22dt}\\
\left<\theta_{2K}(t, {\bf k}_1) \theta_{2K}(t, {\bf k}_2)\right> &=& (2\pi)^3 \delta^{(3)}({\bf k}_1+{\bf k}_2)P_{\theta\theta}^{(22)}(t, k), \label{ps22tt}\\
\left<\delta_{\rm L}({\bf k}_1) \theta_{3K}(t, {\bf k}_2)\right> &=& (2\pi)^3 \delta^{(3)}({\bf k}_1+{\bf k}_2)P_{\theta\theta}^{(13)}(t, k), \label{ps13tt}
\end{eqnarray}
and
\begin{eqnarray}
{1\over 2}\left(\left<\delta_{\rm L}({\bf k}_1) \theta_{3K}(t, {\bf k}_2)\right> + \left<\delta_{3K}(t, {\bf k}_1)\delta_{\rm L}({\bf k}_2) \right>\right) &=& (2\pi)^3 \delta^{(3)}({\bf k}_1+{\bf k}_2)P_{\delta\theta}^{(13)}(t, k). 
\label{2psldt}
\end{eqnarray}

As an example, let us explain the derivation of $P_{\delta\delta}^{(22)}(t, k)$. 
Inserting (\ref{d2k}) into (\ref{ps22dd}), we have
\begin{eqnarray}
\left<\delta_{2K}(t,{\bf k}_1)\delta_{2K}(t,{\bf k}_2)\right>  &=& \left<{1\over (2 \pi)^3}\int d^3 q_1d^3 q_2 \delta^{(3)}({\bf k}_1 - {\bf q}_1 - {\bf q}_2) F_2(t,{\bf q}_1,{\bf q}_2) \delta_{\rm L}({\bf q}_1)\delta_{\rm L}({\bf q}_2)\right.\nonumber\\
&&\left. \hspace{0.2cm}\times {1\over (2 \pi)^3}\int d^3 q_3 d^3 q_4 \delta^{(3)}({\bf k}_2 - {\bf q}_3 - {\bf q}_4) F_2(t,{\bf q}_3,{\bf q}_4)\delta_{\rm L}({\bf q}_3)\delta_{\rm L}({\bf q}_4)\right>
\nonumber\\
&=& {1\over (2 \pi)^6}\int d^3 q_1 d^3 q_3 F_2(t,{\bf q}_1,{\bf k}_1 - {\bf q}_1)F_2(t,{\bf q}_3,{\bf k}_2 - {\bf q}_3) 
\nonumber\\
&& \hspace{4cm} \times \left<\delta_L({\bf q}_1)\delta_L({\bf k}_1 - {\bf q}_1)\delta_L({\bf q}_3)\delta_L({\bf k}_2 - {\bf q}_3)\right>. \label{p221}
\end{eqnarray}
Using the relation that hold for the Gaussian variables, we have
\begin{eqnarray}
\left<\delta_L({\bf q}_1)\delta_L({\bf k}_1 - {\bf q}_1)\delta_L({\bf q}_3)\delta_L({\bf k}_2 - {\bf q}_3)\right> 
&=& 
\left<\delta_L({\bf q}_1)\delta_L({\bf k}_1 - {\bf q}_1)\right> \left<\delta_L({\bf q}_3)\delta_L({\bf k}_2 - {\bf q}_3)\right>
\nonumber\\
&+& \left<\delta_L({\bf q}_1)\delta_L({\bf q}_3)\right> \left<\delta_L({\bf k}_2 - {\bf q}_3)\delta_L({\bf k}_1 - {\bf q}_1)\right>
\nonumber\\
&+& \left<\delta_L({\bf q}_1)\delta_L({\bf k}_2 - {\bf q}_3)\right> \left<\delta_L({\bf k}_1 - {\bf q}_1)\delta_L({\bf q}_3)\right>,
\end{eqnarray}
which yields
\begin{eqnarray}
&&\left<\delta_L({\bf k}_{11})\delta_L({\bf k}_1 - {\bf k}_{11})\delta_L({\bf k}_{21})\delta_L({\bf k}_2 - {\bf k}_{21})\right> \nonumber\\
&&\hspace{4cm}
=
(2\pi)^6 \delta^{(3)}({\bf q}_1 + {\bf q}_3)\delta^{(3)}({\bf k}_1 + {\bf k}_2 - {\bf q}_1 - {\bf q}_3)P_{\rm L}(q_1)P_{\rm L}(|{\bf k}_1 - {\bf q}_1|)
\nonumber\\
&&\hspace{4cm}
+(2\pi)^6 \delta^{(3)}({\bf q}_1+ {\bf k}_2 - {\bf q}_3)\delta^{(3)}({\bf k}_1 - {\bf q}_1 + {\bf q}_3)P_{\rm L}(q_1)P_{\rm L}(q_3),
\end{eqnarray}
with Eq.~(\ref{lps}). Then, (\ref{p221}) yields
\begin{eqnarray}
&&\left<\delta_{2K}(t,{\bf k}_1)\delta_{2K}(t,{\bf k}_2)\right>  
= \delta^{(3)}({\bf k}_1 + {\bf k}_2) \int d^3 q_1 
\Bigl(F_2(t,{\bf q}_1,{\bf k}_1 - {\bf q}_1)F_2(t,- {\bf q}_1, {\bf k}_2 + {\bf q}_1) 
\nonumber\\
&&\hspace{4.5cm} + F_2(t,{\bf q}_1, {\bf k}_1 - {\bf q}_1)F_2(t,{\bf k}_2 + {\bf q}_1,- {\bf q}_1)\Bigr)
P_{\rm L}(q_1)P_{\rm L}(|{\bf k}_1 - {\bf q}_1|).
\end{eqnarray}
Using the relation (\ref{f2ss}), we have
\begin{eqnarray}
P_{\delta\delta}^{(22)}(t,k) = {2\over (2\pi)^3} \int d^3 q_1 F_2^2(t,{\bf q}_1, {\bf k} - {\bf q}_1)P_{\rm L}(q_1)P_{\rm L}(|{\bf k} - {\bf q}_1|),
\end{eqnarray}
which reduces to (\ref{Pdd}). 
In the derivation, we define $x = \cos \theta$, where $\theta$ is the angle 
between ${\bf k}_1$ and ${\bf q}_1$. 
Similarly, the expressions (\ref{Pdt}) and (\ref{Ptt}) are obtained for 
$P_{\delta\theta}^{(22)}(t, k)$ and $P_{\theta\theta}^{(22)}(t, k)$. 
In the limit of the Einstein de Sitter universe withtin the general relativity, 
$\lambda(t) = \lambda_\theta(t) = \mu(t) = \mu_\theta(t) = 1$, which gives the well-known expressions 
\begin{eqnarray}
P_{\delta\delta}^{(22)}(k) &=& {k^3 \over 98(2\pi)^2}\int d r P_{\rm L}(rk)\int^1_{-1} dx P_{\rm L}((1 + r^2 - 2r x)^{1/2})
{(3 r + 7  x -10 r x^2)^2\over (1 + r^2 - 2 r x)^2},\\
P_{\delta\theta}^{(22)}(k)&=& { k^3 \over 98(2\pi)^2}\int d r P_{\rm L}(rk)\int^1_{-1} dx P_{\rm L}((1 + r^2 - 2 r x)^{1/2})\nonumber\\
&&\hspace{4.5cm}\times{(3 r + 7 x -10 r x^2)(- r + 7  x - 6  r x^2)\over (1 + r^2 - 2 r x)^2}\\
P_{\theta\theta}^{(22)}(k) &=& { k^3 \over 98(2\pi)^2}\int d r P_{\rm L}(rk)\int^1_{-1} dx P_{\rm L}((1 + r^2 - 2 r x)^{1/2})
{(- r + 7 x - 6 r x^2)^2\over (1 + r^2 - 2 r x)^2},
\end{eqnarray}
which are constant as functions of time. 

Next, let us explain the derivation of $P_{\delta\delta}^{(13)}(t, k)$. 
Inserting (\ref{d3k}) into (\ref{ps13dd}), we have
\begin{eqnarray}
&&2 \left<\delta_L({\bf k}_1)\delta_{3K}(t,{\bf k}_2)\right> \nonumber\\
&&\hspace{0.7cm} = {2\over (2 \pi)^6}\int d^3 q_1 d^3 q_2 F_3(t,{\bf q}_1, {\bf q}_2, {\bf k}_2 - {\bf q}_1 - {\bf q}_2)
\left<\delta_L({\bf k}_1)\delta_L({\bf q}_1)\delta_L({\bf q}_2)\delta_L({\bf k}_2 - {\bf q}_1 - {\bf q}_2)\right>.
\label{ps311}
\end{eqnarray}
Using the relations,
\begin{eqnarray}
\left<\delta_L({\bf k}_1)\delta_L({\bf q}_1)\delta_L({\bf q}_2)\delta_L({\bf k}_2 - {\bf q}_1 - {\bf q}_2)\right> &=&
(2\pi)^6 \delta^{(3)}({\bf k}_1 + {\bf q}_1) P_{\rm L}(k_1) \delta^{(3)}({\bf k}_2 - {\bf q}_1)P_{\rm L}(q_2)
\nonumber\\
&+&(2\pi)^6  \delta^{(3)}({\bf k}_1 + {\bf q}_2) P_{\rm L}(k_1) \delta^{(3)}({\bf k}_2 - {\bf q}_2)P_{\rm L}(q_1)
\nonumber\\
&+&(2\pi)^6 \delta^{(3)}({\bf k}_1 + {\bf k}_2 - {\bf q}_1 - {\bf q}_2) P_{\rm L}({\bf k}_1) \delta^{(3)}({\bf q}_1 + {\bf q}_2)P_{\rm L}(q_1),
\nonumber\\
\end{eqnarray}
and the symmetries, (\ref{efu3sym}), we have 
\begin{eqnarray}
2P_{\delta\delta}^{(13)}(t,k) = {6\over (2\pi)^3}\int d^3 q_1 F_3(t,{\bf k},{\bf q}_1,- {\bf q}_1) P_{\rm L}(k)P_{\rm L}(q_1).
\end{eqnarray}
After performing the angular integration with respect to 
the spherical coordinate of ${\bf q_1}$, we finally have (\ref{P13MGST}). 
Note that (\ref{P13MGST}) does not depend on $\nu(t)$, which occurs because of the identity 
$\xi({\bf k}, {\bf q}_1, -{\bf q}_1) = 0$. $P_{\delta\delta}^{(13)}(t,k) $ is characterized 
by $\lambda(t)$ and $\mu(t)$. Similarly, we have the expressions (\ref{P13dt})
and (\ref{P13tt}) for $P_{\delta\theta}^{(13)}(t, k)$ and $P_{\theta\theta}^{(13)}(t, k)$, respectively. 
Because of the same reason for $P_{\delta\delta}^{(13)}(t, k)$, 
$P_{\delta\theta}^{(13)}(t, k)$ and $P_{\theta\theta}^{(13)}(t, k)$ do not depend on $\nu(t)$ and 
$\nu_\theta(t)$. Furthermore, $P_{\delta\theta}^{(13)}(t, k)$ and $P_{\theta\theta}^{(13)}(t, k)$ do not depend on $\lambda_\theta(t)$. 
This is because of the nature of the integration 
\begin{eqnarray}
\int dx \hspace{0.1cm}\alpha\gamma_R({\bf k}, {\bf q}_1, -{\bf q}_1) = 0.
\label{zerozero}
\end{eqnarray}
Finally, $P_{\delta\theta}^{(13)}(t, k)$ depends on $\lambda(t)$, $\mu(t)$, and $\mu_\theta(t)$,
and $P_{\theta\theta}^{(13)}(t, k)$ depends on $\lambda(t)$ and $\mu_\theta(t)$. 
We find the following relation holds, in general,
$P_{\delta\theta}^{(13)}(t, k) = [P_{\delta\delta}^{(13)}(t, k) + P_{\theta\theta}^{(13)}(t, k)]/2$, from (\ref{2psldt}).

In the limit of the Einstein de Sitter universe withtin the general relativity, 
all the coefficients $\lambda(t)$, ~$\mu(t)$, ~$\mu_\theta(t)$ reduce to $1$, which 
reproduces the well-known expressions 
\begin{eqnarray}
2P_{\delta\delta}^{(13)}(k) &=& { k^3 \over 252(2\pi)^2 } P_{\rm L}(k)\int dr P_{\rm L}(rk) \nonumber\\
&&\times \left[ 12{1\over r^2} - 158 + 100 {r^2} - 42 {r^4} + {3\over r^3} (r^2 - 1)^3 (7 r^2 + 2) \ln \left({r + 1 \over |r - 1|}\right)\right],\\
2P_{\delta\theta}^{(13)}(k) &=& { k^3 \over252(2\pi)^2 }P_{\rm L}(k)\int d r P_{\rm L}(rk) \nonumber\\
&&\times \left[24 {1\over r^2} - 202 + 56 {r^2} - 30{r^4} + {3\over r^3} \left(r^2 - 1\right)^3 \left(5 r^2 + 4 \right)\ln \left( {r +1\over |r - 1|}\right) \right],\\
2P_{\theta\theta}^{(13)}(k) &=& { k^3 \over 84(2\pi)^2}P_{\rm L}(k)\int d r P_{\rm L}(rk) \nonumber\\
&&\times \left[12 {1\over r^2} - 82 + 4 {r^2} - 6 {r^4} + {3\over r^3} \left(r^2 - 1\right)^3 \left(r^2 + 2 \right)\ln \left( {r + 1\over |r - 1|}\right) \right].
\end{eqnarray}

\section{The integrations of mode-coupling functions}
Here we summarize the useful expressions, which are useful in deriving the 
1-loop order power spectra,
\begin{eqnarray}
\alpha^2({\bf q}_1, {\bf k}_1 - {\bf q}_1) &=& {(r + x - 2 r x^2)^2 \over 4 r^2 (1 + r^2 - 2 r x)^2},\\
\alpha({\bf q}_1, {\bf k}_1 - {\bf q}_1) \gamma({\bf q}_1, {\bf k}_1 - {\bf q}_1) &=& {(r + x - 2 r x^2)(-1 + x^2) \over 2 r (1 + r^2 - 2 r x)^2},\\
\gamma^2({\bf q}_1, {\bf k}_1 - {\bf q}_1) &=& {(- 1 + x^2)^2 \over (1 + r^2 - 2 r x)^2},
\end{eqnarray}
and 
\begin{eqnarray}
&&\int d^3 q_1 \hspace{0.1cm} P_{\rm L}(rk) \alpha\alpha({\bf k}, {\bf q}_1, - {\bf q}_1) = {2 \pi k^3 \over 72}  \int d r P_{\rm L}(rk)\left[- 2 + 16 r^2 - 6 r^4 + 
{3\over r^3}(r^2 - 1)^3\ln \left({r + 1\over \left| r - 1\right|}\right)\right],\\
&&\int d^3 q_1 \hspace{0.1cm} P_{\rm L}(rk) \alpha\gamma_R({\bf k}, {\bf q}_1, - {\bf q}_1) = 0, \\
&&\int d^3 q_1 \hspace{0.1cm} P_{\rm L}(rk) \alpha\gamma_L({\bf k}, {\bf q}_1, - {\bf q}_1) = {2 \pi k_1^3 \over 36} \int d r P_{\rm L}(rk)\left[6 + 16 r^2 - 6r^4 + 
{3\over r^3}(r^2 - 1)^3\ln \left({r + 1\over \left| r - 1\right|}\right)\right],\\
&&\int d^3 q_1 \hspace{0.1cm} P_{\rm L}(rk) \gamma\gamma({\bf k}_1, {\bf q}_1, - {\bf q}_1) = {2 \pi k_1^3\over 72} \int d r P_{\rm L}(rk)\left[- 6 {1 \over r^2} + 22 + 22 r^2 - 6 r^4  + {3\over r^3}(r^2 - 1)^4 \ln \left({r + 1\over \left| r - 1\right|}\right)\right], \nonumber\\ \\
&&\int d^3 q_1 \hspace{0.1cm} P_{\rm L}(rk) \xi({\bf k}_1, {\bf q}_1, - {\bf q}_1) = 0.
\end{eqnarray}

\section{Coefficients and  in the KGB model}
In the KGB model, we find the coefficients in basic equations, 
\begin{eqnarray}
&&{\cal F}_T = M_{\rm pl}^2,~~~~~{\cal G}_T =M_{\rm pl}^2,\\
&&\Theta = - n M_{\rm pl}\left({r_c^2 \over M_{\rm pl}^2}\right)^n \dot{\phi} X^n + H M_{\rm pl}^2,\\
&&\dot{\Theta} = -n(2n + 1)M_{\rm pl}\left({r_c^2 \over M_{\rm pl}^2}\right)^n \ddot{\phi} X^n + \dot{H}M_{\rm pl}^2,\\
&&{\cal E}= - X + 6 n M_{\rm pl}\left({r_c^2 \over M_{\rm pl}^2}\right)^n \dot{\phi}H X^n - 3 H^2 M_{\rm pl}^2,\\
&&{\cal P}= - X - 2n M_{\rm pl}\left({r_c^2 \over M_{\rm pl}^2}\right)^n\ddot{\phi}X^n + (3 H^2 + 2\dot{H})M_{\rm pl}^2,\\
&&A_0 = {X\over H^2} - 2nM_{\rm pl}\left({r_c^2 \over M_{\rm pl}^2}\right)^n \left({2\dot{\phi}\over H} + n{\ddot {\phi}\over H^2}\right)X^n,\\
&&A_2 = B_0 = n {\dot{\phi}\over H}M_{\rm pl}\left({r_c^2 \over M_{\rm pl}^2}\right)^n X^n,\\
&&A_1 = B_1 = B_2 = B_3 = C_0 = C_1 = 0, 
\end{eqnarray}
and the non-trivial expressions,
\begin{eqnarray}
&&L(t)=-{A_0 {\cal F}_T \rho_{\rm m}\over 2(A_0 {\cal G}_T + A_2^2 {\cal F}_T)},\\
&&N_{\gamma}(t)={B_0 A_2^3 {\cal F}_T^3 \rho_{\rm m}^2 \over 4(A_0 {\cal G}_T^2 + A_2^2 {\cal F}_T)^3H^2},
\\
&&H^2 \mu_\Phi = - {8 B_0 {\cal T}^3 \rho_{\rm m}^2 \over 7 H^2 {\cal Z}^3}\lambda  - {8 B_0^2 {\cal G}_T^2 {\cal T}^4 \rho_{\rm m}^3 \over H^4 {\cal Z}^5}. 
\end{eqnarray}
We use the attractor solution which satisfies  $3\dot\phi H G_{3X}=1$. Then we have 
\begin{eqnarray}
&&\ddot{\phi} = - {1 \over 2n-1}{\dot{\phi} \dot{H}\over H},\\
&&{\dot{H} \over H^2} = - {(2n-1) 3 \Omega_{\rm m} \over 2(2n - \Omega_{\rm m})},
\\
&&A_0 
=-{M_{\rm pl}^2(1-\Omega_{\rm m})\left(2n +(3n - 1)\Omega_{\rm m}\right) 
\over 2n-\Omega_{\rm m}},\\
&&A_2=
M_{\rm pl}^2{(1-\Omega_{\rm m})},\\
&&B_0=
M_{\rm pl}^2{(1-\Omega_{\rm m})},
\end{eqnarray}
where we define $\Omega_{\rm m}=\rho_m(a)/3M_{\rm pl}^2H^2$. We also have
\begin{eqnarray}
{\cal R} &=& - {M_{\rm pl}^4(1-\Omega_{\rm m})\left(2n +(3n - 1)\Omega_{\rm m}\right)\over 2n-\Omega_{\rm m}},\\
{\cal S} &=& - {M_{\rm pl}^4(1-\Omega_{\rm m})\left(2n +(3n - 1)\Omega_{\rm m}\right)\over 2n-\Omega_{\rm m}},\\
{\cal T} &=& M_{\rm pl}^4(1-\Omega_{\rm m}),\\
{\cal Z} &=& 2{M_{\rm pl}^6 \Omega_{\rm m}(5n - \Omega_{\rm m})(1-\Omega_{\rm m})\over 2n - \Omega_{\rm m}}.
\end{eqnarray}


\end{document}